\documentclass[fleqn,usenatbib]{mnras}

\usepackage{newtxtext,newtxmath}

\usepackage[T1]{fontenc}

\DeclareRobustCommand{\VAN}[3]{#2}
\let\VANthebibliography\thebibliography
\def\thebibliography{\DeclareRobustCommand{\VAN}[3]{##3}\VANthebibliography}

\usepackage{graphicx}	
\usepackage{amsmath}	
\usepackage{pifont}
\usepackage{mathtools}
\usepackage{placeins}
\usepackage{float}
\usepackage{bm}
    
\usepackage{nccmath}
\newcommand\at[2]{\left.#1\right|_{#2}}
\usepackage{cleveref}
\crefname{figure}{Fig.}{Figs.}
\crefname{table}{Table}{Tables}

\renewcommand{\vec}[1]{ {\bf #1} }

\title[Simulating cold shear flows on a moving mesh]{Simulating cold shear flows on a moving mesh}

\author[O. Zier and V. Springel]{%
Oliver Zier$^{1}$\thanks{E-mail: ozier@mpa-garching.mpg.de}
and Volker Springel$^{1}$
\\%
$^{1}$Max-Planck-Institut für Astrophysik, Karl-Schwarzschild-Straße 1, 85741 Garching, Germany\\
}

\date{Accepted XXX. Received YYY; in original form ZZZ}

\pubyear{2022}

\begin{document}
\label{firstpage}
\pagerange{\pageref{firstpage}--\pageref{lastpage}}
\maketitle

\begin{abstract}
Rotationally supported, cold, gaseous disks are ubiquitous in astrophysics and appear in a diverse set of systems, such  as protoplanetary disks, accretion disks around black holes, or large spiral galaxies.  Capturing the gas dynamics accurately in these systems is challenging in numerical simulations due to the low sound speed compared to the bulk velocity of the gas, the resolution limitations of full disk models, and the fact that numerical noise can easily source spurious growth of fluid instabilities if not suppressed sufficiently well, negatively interfering with real physical instabilities present in such disks (like the magneto-rotational instability). Here we implement the so-called shearing-box approximation in the moving-mesh code {\small AREPO} in order to facilitate achieving  high resolution in local regions of differentially rotating disks and to address these problems. While our new approach offers manifest translational invariance across the shearing-box boundaries and offers continuous local adaptivity, we demonstrate that the unstructured mesh of {\small AREPO} introduces unwanted levels of ``grid-noise'' in the default version of the code. We show that this can be rectified by high-order integrations of the flux over mesh boundaries. With our new techniques we obtain highly accurate results for shearing-box calculations of the magneto-rotational instability that are superior to other Lagrangian techniques. These improvements are also of value for other applications of the code that feature strong shear flows.
\end{abstract}


\section{Introduction}

Computer simulations are one of the main methods to understand the nonlinear evolution of gaseous flows in accretion disks and other differentially rotating systems. However, following key local instabilities such as the magnetorotational instability (MRI) \citep{velikhov1959stability, chandrasekhar1960stability, fricke1969stability, balbus1991powerful} requires high resolution, which makes global simulations overly expensive. But the MRI is by no means the only instability of interest. For example, in protoplanetary disks, the Rossby wave instability appears \citep{Lovelace1999}, and radially stratified disks give rise to a variety of further instabilities that may drive turbulence in the disk \citep[e.g.][]{Klahr2014}. In the context of the interstellar medium, differential shear is also considered an important source of turbulence, but one that is challenging to capture accurately in global disk models as a result of the limited resolution available in such calculations. 

A general difficulty in the numerical treatment of rotationally supported gas flows lies in the ``coldness'' of the gas relative to the rotational bulk motion. The gas velocities in the rest frame of the disk are dominated by the coherent rotational motion and can be far higher than the involved sound speeds, making the flow formally highly supersonic. This causes small timesteps and large advection errors in standard finite-volume hydrodynamical grid codes. On the other hand, standard Lagrangian methods such as smoothed particle hydrodynamics (SPH)  are prone to seed spurious instabilities in the disks  due to their inherent, comparatively large numerical noise.

An interesting alternative lies in the use of a local approximation \citep{hill1878collected,goldreich1965ii} as realized in the shearing box approach, that was first applied in \citet{hawley1995local} to the MRI. The equations describing the evolution of the system are here expressed in a local, Cartesian reference frame that is co-rotating with the local angular velocity $\Omega_0 = \Omega(r_0)$ at a radius $r_0$ in the disk. This is typically combined with imposing periodic boundaries in the azimuthal direction (i.e. one overs only a region $\Delta y = r_0 \Delta \phi$), and a special `shear-periodic' boundary condition in the radial direction (i.e. one only covers a radial region $\Delta r$). The transformation to a non-inertial frame of reference leads  to two new source terms compared to the standard ideal magnetohydrodynamical (MHD) equations: The Coriolis force and the centrifugal force. The latter can also be written as  an effective gravity force, defined as the difference of the external gravitational force and the centrifugal force. This effective forces cancels for an equilibrium solution with a linear shear flow in the azimuthal direction, corresponding to a plain differential rotation in the disk. Fluid instabilities around this ``cold ground state'' can then be studied with high accuracy with Eulerian methods.

The shearing box approximation is only valid as long as the size of the used box, $L_r = \Delta r$, is much smaller than $r_0$. As a result, it does not readily allow for measurements of global disk properties such as the net mass accretion rates, or radial density and temperature profiles \citep{regev2008viability}. Despite these limitations, it has become a very important standard method to analyze the nonlinear saturation of the MRI \citep{hawley1995local}. In particular, due to its much lower computational cost compared to global disk models, it allows the exploration of large parameter spaces, and thus ultimately yields important predictions for full simulations as well. For example, \cite{riols2018magnetorotational, riols2019gravitoturbulent} found that gravoturbulence can act as a dynamo and amplify MHD fields even in the presence of strong resistivity, a result later confirmed by \citet{deng2020global} using global simulations.

So far simulations with the shearing box approximation have mostly employed Eulerian codes that require a special mapping of the fluid states at the radial boundary conditions. Examples for such implementations are the finite volume codes {\small ATHENA} \citep{stone2008athena, stone2010implementation}, {\small PLUTO} \citep{mignone2007pluto} and {\small RAMSES} \citep{fromang2006high}, or the finite difference codes {\small PENCIL} \citep{brandenburg2020pencil} and {\small ZEUS} \citep{stone1992zeusa, stone1992zeusb, hawley1995local}.

Although Lagrangian methods have often been used for global disk simulations due to their lower advection errors, they have only rarely been used for shearing box simulations,  despite the fact that they in principal  allow for a translationally invariant implementation of the radial boundary conditions. One reason is that particle-based methods have problems sustaining MRI turbulence without a net magnetic field, as shown by \cite{deng2019local} with the {\small GIZMO} code \citep{hopkins2015new}. However, \citet{wissing2022magnetorotational} demonstrated recently that some SPH implementations are able to sustain the turbulence after all, which is a prerequisite for more complicated shearing box studies.

In this work, we present the first implementation of the shearing box approximation in a moving-mesh code, specifically the  {\small AREPO}  code \citep{springel2010pur, weinberger2020arepo}, which can be viewed as  a hybrid method between a static grid code and a Lagrangian particle method. A full implementation of the shearing box approximation consists of the source terms and a treatment of the special radial boundary conditions. In contrast to Eulerian codes, the moving-mesh approach allows a fully translationally invariant implementation of the radial boundary conditions at the discretized level of the equations, so that the radial boundaries are manifestly indistinguishable from interior regions of the flow. Another advantage is that the moving-mesh approach retains its ability to continuously and seamlessly adjust the local resolution, making it particularly suitable for calculations of fragmentation and local gravitational collapse.  Interestingly, our results for the ground-state of the shearing-box show that the default version of {\small AREPO} produces comparatively large `grid-noise' errors when the mesh becomes geometrically completely unstructured due to strong local shear. We found that these errors can be eliminated, however, with a novel high-order flux integration that we develop as part of this paper.

This paper is structured as follows: In Section~\ref{sec:OverViewArepo} we summarize the most important ingredients of the moving mesh method as currently implemented in {\small AREPO}. In Section~\ref{sec:implementationShearingBox} we present our implementation of the source terms and boundary conditions for the shearing box. We show in Section~\ref{sec:improvedArepo} that the vanilla version of {\small AREPO} leads to significant noise in the ground state, discuss several improvements that reduce this noise and show that they can also improve the results in other problems such as the simulation of a full cold Keplerian disk.  In Section~\ref{sec:testProblems} we run several test problems for the shearing box and show that our implementation converges with close to second-order, as expected. Finally, we show in Section~\ref{sec:nonlinearMRI} that {\small AREPO} is able to sustain MRI turbulence for several hundred orbits without a background magnetic field and compare our results to those from \citet{shi2016saturation} obtained with the {\small ATHENA} code. In Section~\ref{sec:discussion} we summarize our results and present our conclusions.

\section{Ideal magnetohydrodynamics in Arepo}

\label{sec:OverViewArepo}
The moving-mesh code {\small AREPO} \citep{springel2010pur, weinberger2020arepo} solves the ideal magnetohydrodynamical (MHD) equations on an unstructured Voronoi mesh using the finite volume method \citep{pakmor2011magnetohydrodynamics, pakmor2013simulations}. The computational mesh is constructed using mesh-generating points that can move with an arbitrary velocity, but which is typically close to the fluid velocity. Due to this movement, the Voronoi mesh changes its structure and topology in time and the method becomes manifestly Galilei-invariant when the mesh motion is tied to the fluid motion itself.

The ideal MHD equations in differential form can be expressed as:
\begin{equation}
    \frac{\partial \bm U}{\partial t} + \nabla \cdot \bm F(\bm U) = 0,
    \label{eq:idealMHDDifferential}
\end{equation}
which exposes the hyperbolic conservation law for the mass, momentum, energy and integrated magnetic field.
$\bm U$ defines here the conserved quantities and $\bm F$ is the flux function. The flux function is defined in a local rest frame by:
\begin{equation}
\bm U  = \begin{pmatrix}
   \rho \\
   \rho \bm v  \\
   \rho e  \\
   \bm B\\
   \end{pmatrix}, \;\;\;\;\;\;
   F(\bm U) =  \begin{pmatrix}
   \rho \bm v\\
   \rho \bm v \bm v^T + P -\bm B \bm B^T\\
   \rho e  \bm v + P \bm v -\bm B(\bm v \cdot \bm B)\\
   \bm B \bm v^T -\bm v \bm B^T,
   \end{pmatrix},
   \label{eq:idealMHDFlux}
\end{equation}
where $\rho$, $\bm v$, $e$, $\bm B$, $P$ are the density, velocity, total energy per mass, magnetic field strength and pressure, respectively. The total energy density $e = u + \frac{1}{2} \bm v^2 + \frac{1}{2 \rho} \bm B^2$ consists of the thermal energy  per mass $u$, the kinetic energy density $\frac{1}{2} \bm v^2$ and the magnetic field energy density $\frac{1}{2 \rho} \bm B^2$. The pressure $P=p_{\rm gas} + \frac{1}{2}\bm B^2$ has a thermal and a magnetic component. The system of equations is closed by the equation of state (EOS), that expresses $p_{\rm gas}$ as a function of the other thermodynamical quantities. We shall adopt the usual ideal gas EOS, $p_{\rm gas} = (\gamma -1) \rho u$, with an adiabatic index $\gamma$.

The ideal MHD equations fulfill the condition $\nabla \cdot \bm B = 0$ for all times if the initial field is divergence free. This property of the continuum equations is unfortunately lost in discretized versions of the equations, unless elaborate constrained transport formulations are used that can retain the divergence free conditions also at the discrete level. In this paper, we will instead use the Dedner scheme \citep{dedner2002hyperbolic} for divergence control as implemented in \cite{pakmor2011magnetohydrodynamics}. It diffuses away local deviations of the divergence constraint and damps them. We note that in many studies with {\small AREPO} the Powell scheme \citep{powell1999solution} as implemented in \cite{pakmor2013simulations} is used instead, but all results except those in Section~\ref{sec:nonlinearMRI} are insensitive to the details of the divergence cleaning. We therefore defer a comparison between the two methods in the context of the shearing box to a future paper. We set the parameter $c_h$ required in the Dedner approach to the maximum signal speed in the system, and choose $c_p=\sqrt{2 c_h r}$ in proportion to the cell radius $r$ (see \cite{pakmor2011magnetohydrodynamics} for details).

The ideal MHD equations are discretized on the cells of the dynamic Voronoi mesh. Volume averages of the conserved quantities for a cell $i$ can be calculated by integrating $\bm U$ over the corresponding volume $V_i$:
\begin{equation}
   \bm Q_i = \int_{V_i} \bm U.
\end{equation}
By integrating equation (\ref{eq:idealMHDDifferential}) over $V_i$ and applying Gauss's law we find for the change of $\bm Q_i$:
\begin{equation}
    \frac{{\rm d} \bm Q_i}{{\rm d} t} = \int_{\partial V_i} \bm F(\bm U) \cdot \hat{n}.
\end{equation}
$\hat{n}$ is the outward normal on the surface $\partial V_i$ of cell $i$. The temporal evolution of $\bm Q_i$ over one time step of size $\Delta t$ can therefore be written as:
\begin{equation}
 \bm Q_i^{n+1} =\bm Q_i^n +  \int_{t_0}^{t_0 + \Delta  t} {\rm d}t  \int_{\partial V_i} \bm F(\bm U) \cdot \bm \hat{\bm n}.
 \label{eq:TimeIntegrationGeneral}
\end{equation}
The time integral is discretized in {\rm AREPO}  using a second-order accurate hybrid  between a Runge–Kutta method and the MUSCL-Hancock scheme \citep[see][]{pakmor2016improving}. By defining $\bm I(t) = \int_{\partial V_i} \bm F(\bm U) \cdot \bm \hat{\bm n}$,  equation (\ref{eq:TimeIntegrationGeneral}) can be approximated by:
\begin{equation}
    \bm Q_i^{n+1} = \bm Q_i^n +\frac{\Delta t}{2} \left[\bm I(t_0) + \bm I(t_0+ \Delta t)\right].
\end{equation}

The integral $\bm I(t)$ can be expressed as a sum of integrals $\bm I_{ij}$ over all outer faces of the cell. Such a face integral is then approximated as
\begin{equation}
    \bm I_{ij}(t) =\int_{A_{ij}} \bm F(\bm U) \cdot \hat{\bm n}_{ij} dA_{ij}  \approx A_{ij} \bm F[\bm U(\bm f)] \cdot \hat{\bm n}_{ij},
    \label{eq:approxIntegralMid}
\end{equation}
where $ \bm U(\bm f)$ is the state at the geometric centre $\bm f$ of the face, and $A_{ij}$ is the interface between cell $i$ and $j$. 

The state at the interface is taken as the solution of a Riemann problem with the (linearly extrapolated) state of the cells $i$ and $j$ at this point as input. These extrapolations are calculated based on a slope-limited piece-wise linear spatial reconstruction step in each cell. For the calculation of $\bm I(t+\Delta t)$, an additional first-order time extrapolation of the fluid states by $\Delta t$ is applied. This reconstruction is done using primitive variables for expressing the fluid state:
\begin{equation}
    \bm W = \begin{pmatrix}
   \rho \\
    \bm v  \\
  p_{\rm gas}  \\
   \bm B\\
   \end{pmatrix},
\end{equation}
which can be easily obtained from the conserved quantities.  

The gradients of the primitive variables are estimated using a least-square fit around every cell \citep[see][for details]{pakmor2016improving} and the reconstruction takes the following form for the left and right state:
\begin{equation}
    \bm W_{L,R}' (t_0 + \Delta t) = \bm W_{L,R} (t_0)+ \at{\frac{\partial \bm W}{\partial \bm r}}{L,R} (\bm f - \bm s_{L,R}) + \at{\frac{\partial \bm W}{\partial t}}{L,R} \Delta t ,
    \label{eq:reconstructionGeneral}
\end{equation}
where $\bm s$ is the center of mass of each cell. $\frac{\partial \bm W}{\partial t}$ is only required in the second flux calculation and can be obtained from the linearized ideal MHD equations:
\begin{equation}
    \frac{\partial}{\partial t} \begin{pmatrix}
   \rho \\
    \bm v  \\
  p_{gas}  \\
   \bm B\\
   \end{pmatrix} = \begin{pmatrix}
   -\bm v \nabla \rho - \rho \nabla v \\
   - \frac{\nabla P}{\rho} - \bm v \bm v^T + \bm B \cdot (\nabla \bm B)  \\
    - \gamma p_{gas} \nabla \cdot\bm v - \bm v \cdot\nabla P\\
   \bm B \cdot (\nabla \bm v) - (\nabla \cdot \bm v) \bm B - \bm v \cdot (\nabla \bm B)\\
   \end{pmatrix}.
   \label{eq:linearTimeExtrapolation}
\end{equation}
As detailed in \cite{pakmor2011magnetohydrodynamics} we first try, in consecutive order, the HLLD Riemann solver \citep{miyoshi2005multi}, the HLL solver \citep{harten1983upstream}, and in case both fail, the Rusanov solver \citep{rusanov1961calculation}.

If the mesh itself is moving with velocity $\bm v_{m}$ one has to add the term $\bm U \bm v_{m}^T$ to the flux function. The velocity of a point on the interface can then be calculated as in \citet[][their eqn.~33]{springel2010pur}, to which we refer for more details. It only requires the local geometry of the two neighbouring cells of the interface as well as their velocities. The motion of the mesh generating points is typically kept as close as possible to the velocity of the fluid, but {\small AREPO} also adds small additional velocity corrections to improve the local regularity of the Voronoi mesh. We will refer to the latter as mesh regularisation in the following. As measure for the local mesh quality we
use the maximum angle under which any cell face is seen by a mesh generating point, as described in full in  \cite{vogelsberger2012moving, weinberger2020arepo}. Unless specified otherwise, 
we use the parameters $\beta =2$ and $f_{\rm shaping} = 0.5$ as defaults for this method.

\begin{figure*}
    \centering
    \includegraphics[width=1\linewidth]{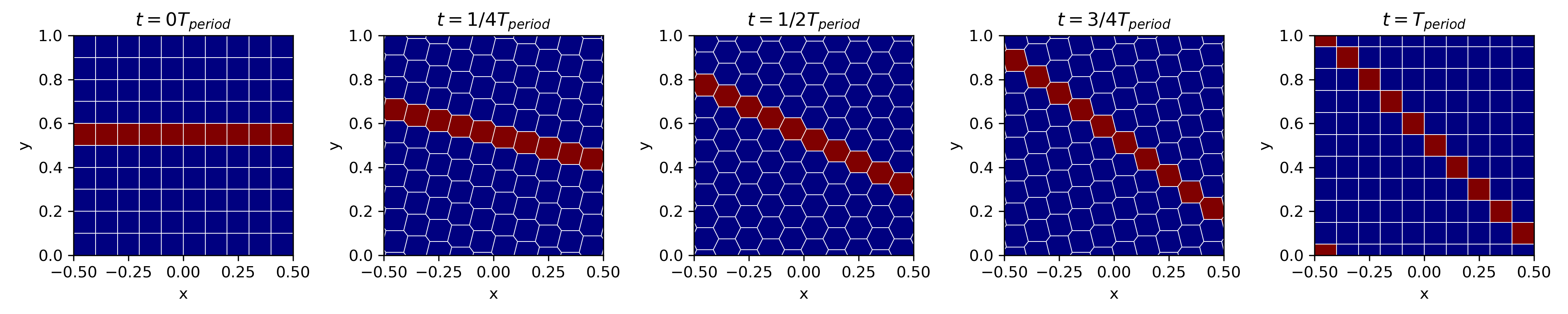}
    \caption{The shearing of an initially Cartesian grid by the background flow (\ref{eq:groundState}). We define $T_{\rm period} = \Omega_0 L_y / (q L_x) $ as the time it takes to evolve the grid into a new Cartesian grid. During this time the geometry of the grid continuously changes while retaining a residual symmetry of the cells (see, e.g., the red colored cells).}
    \label{fig:cartoonShearingCartesianGrid}
\end{figure*}

\begin{figure}
    \centering
    \includegraphics[width=0.95\linewidth]{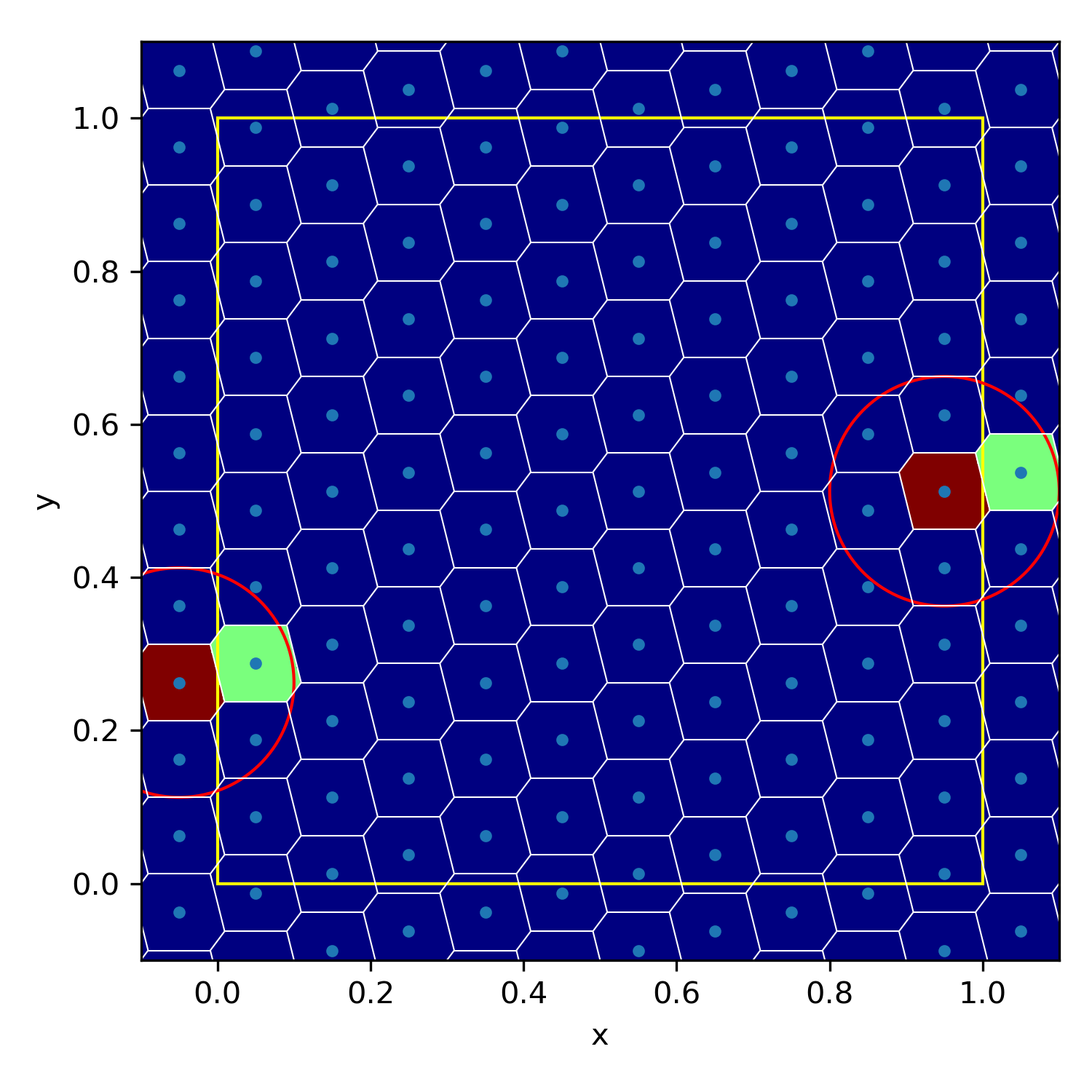}
    \caption{Illustration of the mesh construction for the shearing box boundary conditions in the $x$-direction: The inside of the yellow square represents the two-dimensional simulation domain. All cells whose mesh generating points (blue circles) lie in the box are primary cells, the other ones are so-called ghost cells arising as duplicates due to the boundary conditions. During the construction of the mesh, we need to repeatedly find all mesh-generating points that lie within a sphere of some radius $r$ around a mesh-generating point. These are in-circle tests for the circumspheres of triangles, carried out to make sure that the corresponding triangle is a valid Delaunay triangle of the tessellation (see as example the red circle around a cell colored in red). If this sphere intersects an $x$-boundary, we have to calculate an additional search position by shifting the $x$-coordinate by $\pm L_x$ and the y-coordinate by $\mp wt$, see eqn.~(\ref{eq:shearingBoxBoundaryConditions}). For this new position, another in-circle test has to be carried out, which may in turn return both primary points or other shifted ghost points. Note that in our example, the interface between the red and the green cell exists twice in the final tessellation: Once on the left hand side with the red cell as ghost cell, and once on the right side with the green one as ghost cell.  It is however sufficient to calculate the corresponding fluxes, together with the boost of eqn.~(\ref{eq:fluxBoostBoundary}), only once.
    }
    \label{fig:illustration_boundary_conditions}
\end{figure}

\section{Implementation of the shearing box approximation}

\label{sec:implementationShearingBox}
The local shearing box approximation \citep{goldreich1965ii} solves the ideal MHD equations in a frame that is rotating with a frequency $\Omega_0 = \Omega(r_0)$  at radius $r_0$. Upon transformation into the non-inertial rotating system, the equations can be written in Cartesian coordinates $x$, $y$, and $z$, with corresponding units vectors $\bm \hat{e}_x$, $\bm \hat{e}_y$ and $\bm \hat{e}_z$, as:
\begin{equation}
    \frac{\partial \bm U }{\partial t}+ \nabla \cdot \bm F(\bm U) = \bm S_{\rm grav} + \bm S_{\rm cor}.
    \label{eq:shearingBoxEquation}
\end{equation}
The left hand side is formally equivalent to equation~(\ref{eq:idealMHDDifferential}), and the flux is still given by (\ref{eq:idealMHDFlux}), but the coordinate transformation leads to two additional source terms for the momentum and energy equations. $\bm S_{\rm grav}$ contains one of the terms, the centrifugal force, as well as  
the external gravitational force expanded to first order in the $x$- and $z$-directions,  while $\bm S_{\rm cor}$ represents the Coriolis force. Explicitly, 
they are given by:
\begin{equation}
\bm S_{\rm grav}  = \begin{pmatrix}
   0 \\
   \rho \Omega_0^2 \left(2q x \bm\hat{e}_x - z \bm\hat{e}_z\right)  \\
    \rho \Omega_0^2 \bm v \cdot \left(2q x \bm\hat{e}_x - z  \bm\hat{e}_z\right)   \\
   0\\
   \end{pmatrix} 
\end{equation}
and
   \begin{equation}
   \bm S_{\rm cor} =  \begin{pmatrix}
   0\\
   -2 \rho \Omega_0  \bm\hat{e}_z \times \bm v\\
  0\\
  0,
   \end{pmatrix}.
   \label{eq:shearingBoxSourceTerms}
\end{equation}
The term $\bm S_{\rm grav}$ depends on the shearing parameter
\begin{equation}
    q = - \frac{1}{2} \frac{d \ln \Omega^2}{d \ln r},
\end{equation}
which takes on the value $q=3/2$ for a Keplerian rotation curve.

In the azimuthal $y$-direction one typically uses standard periodic boundary conditions. In the $z$-direction, one can use periodic boundary conditions in the case of an unstratified box (no gravitational term in the $z$-direction) or inflow-outflow boundary conditions for a stratified box (gravitational field towards the $z=0$ plane). The boundary conditions in the radial direction are more complicated and will be discussed separately below.

We note that equation (\ref{eq:shearingBoxEquation}) admits a `ground state' solution with
\begin{equation}
    \bm v_0(x) = (0,-q \Omega_0 x,0), \;\;\; \rho = {\rm const.},  \;\;\;  P = {\rm const.},
    \label{eq:groundState}
\end{equation}
in the unstratified case. For a global disk simulation, this is equivalent to the intrinsic radial dependence of the rotational velocity being given by $r\Omega(r)$ everywhere.
This background flow alone leads to a continuous shearing between neighbouring cells in the $x$-direction (see \cref{fig:cartoonShearingCartesianGrid}).

\subsection{Implementation of the source terms}
\label{subsec:sourceTermsImplementation}

To implement the source terms we use a second-order accurate Strang splitting scheme, which means we first evolve the system for  half a time step using the source terms, then for a full time step using the methods from Section~\ref{sec:OverViewArepo}, and then apply another half step for the source terms. We apply both source terms together and actually rewrite the momentum equation to:
\begin{equation}
    \bm S_{\rm tot, mom} = \begin{pmatrix}
   2 (v_y - v_{0,y})\rho \Omega_0 \\
   -2 v_x \rho \Omega_0  \\
  - \rho \Omega^2 z\\
   \end{pmatrix},
   \label{eq:shearingBoxsourceTermImplementation}
\end{equation}
because this term vanishes for the ground state solution. To conserve the total energy accurately we first subtract the kinetic energy from the total energy, apply the momentum source term and then add the new kinetic energy to the total energy.
To avoid an unphysical growth of the amplitude of epicycle oscillations (see section \ref{sec:epicycleOsci}), we add a prediction step for the second half step of the source terms. In practice, this means we first apply the source term (\ref{eq:shearingBoxsourceTermImplementation}) for a half time step, calculate the new velocity and use this velocity for the actual update of the momentum. This method is similar to the implementation the of Crank-Nicholson time difference method in \cite{stone2010implementation}, which actually can conserve the amplitude to machine precision.

\subsection{Boundary conditions}

While we can use the standard implementation of periodic or inflow/outflow boundary conditions in the $y$- and $z$-directions, we have to take care of the background shear flow (\ref{eq:groundState}) in the $x$-direction.
This leads to the so-called shearing box boundary conditions:
\begin{equation}
    f(x,y,z,t) = f(x \pm L_x, y \mp w t, z,t); f= (\rho, \rho v_x, \rho v_z, \bm B),
\end{equation}
\begin{equation}
    \rho v_y(x,y,z,t) = \rho v_y(x \pm L_x, y \mp w t, z,t) \mp \rho  w ,
\end{equation}
\begin{equation}
    e(x,y,z,t) = e(x \pm L_x, y \mp w t, z,t) \mp \rho v_y v_w + \frac{\rho w^2}{2},
\end{equation}
\label{eq:shearingBoxBoundaryConditions}
where $L_x$ is the box size in the (radial) $x$-direction, and $w \equiv q \Omega_0 L_x$. 

To implement standard periodic boundary conditions, {\small AREPO} uses ghost cells. During the mesh construction, all cells closer than a specific radius $r$ to a cell have to be found. If this sphere of influence intersects a box boundary, the search will be continued on the opposite boundary of the box. The cells found there will be built into the local Voronoi mesh as a ghost cell at the original position of the cell shifted by one box length, and the properties of the original cell are copied to the ghost cell. Updates of the conserved quantities of such ghost cells are applied to the original cell \citep[for details of the mesh construction see][]{springel2010pur}. The creation of ghost cells may cause some cell interfaces  to be duplicated in the constructed tessellation (in particular also if parallelization subdivides the fiducial global mesh into several logical pieces stored on different processors), but the fluxes are calculated always only once, such that the conserved quantities of both involved cells get updated only once in a manifestly conservative fashion.

The radial shearing box boundary conditions can be treated very similarly with some extra modifications. First, if the search sphere overlaps with the box boundary, the centre of the search region does not only have to be shifted by $\pm L_x$ in the $x$-direction, but also by $\mp w t$ in the $y$-direction. When copying the properties of the original cell to the ghost cells, we have to add a velocity boost $\mp w$ to the $y$-component of the velocity, as well as to the velocity of the corresponding mesh generating point. The Riemann solver then returns the state $\bm U(\rho, \bm v, p,\bm B)$, which can then be used to calculate the flux $\bm F(\bm U)$ that is applicable to a real (unboosted) cell at the interface. If we are dealing with a ghost cell instead, the conserved quantities of the corresponding original cell need to updated with 
 a modified flux:
\begin{equation}
    \Delta \bm F = \bm F' (\bm U'(\rho, \bm v \pm w \hat{e}_y, p,\bm B)) \mp \bm U'(\rho, \bm v \pm w \hat{e}_y, \bm B) (w \bm \hat{e}_y)^T - \bm F(\bm U).
\end{equation}
$\bm U'$ is here the state we obtain by adding the velocity shift $ \pm w \hat{e}_y$ to $\bm U$.
Using equation (\ref{eq:idealMHDFlux}) for the flux,  this can explicitly be expressed as:
\begin{equation}
    \Delta \bm F  = \begin{pmatrix}0\\\ (0,\pm w Q_{1},0)\\ \pm Q_{2,y} w + 1/2 Q_1 w^2\\ (0,\mp w (\bm B \cdot \hat{n}),0)\end{pmatrix}
\end{equation}
with 
\begin{equation}
 \bm F(\bm U) =  \begin{pmatrix}Q_1\\ (Q_{2,x},Q_{2,y},Q_{2,z})\\Q_3\\(Q_{4,x}, Q_{4,y}, Q_{4,z}) \end{pmatrix},
    \label{eq:fluxBoostBoundary}
\end{equation}
where $\hat{n}$ is the normal of the interface.

Note that this boost shows that in the shearing box approximation the momentum and energy are not conserved. Whereas the mean radial and vertical magnetic field,
\begin{equation}
    \left< B_{x,z} \right> = \frac{\int_{\rm Box} B_{x,z} {\rm d}V}{\int_{\rm Box}  {\rm d}V} = {\rm const.}
\end{equation}
are conserved, the azimuthal field grows linear with the net radial magnetic field flux through the radial boundary \citep{gressel2007shearingbox}:
\begin{equation}
  \frac{\partial \left < B_y\right>}{\partial t}  = -\frac{w}{V}\int_{\partial x} B_x\, {\rm d}y \,{\rm d}z.
\end{equation}
For $\nabla \cdot \mathbf{B} = 0$, the last conditions also implies that $\left < B_y\right>$ is constant if there is no net radial field.

On a moving mesh, cells can also leave the simulation box and reappear on the opposite side. If this happens in the $x$-direction, we also have to modify the $y$-position and velocity component of the cell, as well as its energy. \cref{fig:illustration_boundary_conditions} further illustrates the implementation of the boundary conditions during the mesh construction. We note that this implementation of the boundary conditions does not introduce any special treatment of the boundaries at the level of the flux calculation, and therefore the whole simulation box stays isotropic and translationally invariant.

\section{Reducing numerical noise in Arepo}
\label{sec:improvedArepo}

In this section, we first test the accuracy of the shearing box implementation in {\small AREPO} when 
the default integration method of the code is used. As this exposes an unwanted level of `grid-noise' that can influence the growth of instabilities and act as a numerical viscosity, we then introduce 
several improvements to the code that substantially increase the achieved accuracy.

\subsection{Simulations of the ground state of the shearing box}
\label{subsec:groundStateSimulationOld}

A conceptionally simple but nevertheless highly illuminating test case of the basic code performance lies in simulating the ground state (\ref{eq:groundState}) of the shearing box with an isothermal equation of state. This state should be stable (because the shearing motion is stabilized by the centrifugal force), and indeed, static grid codes are able to maintain this state to machine precision on a Cartesian grid. However, the free mesh motion of {\small AREPO} allows for the occurrence of diverse cell geometries, potentially introducing much larger deviations from the ground state.

\begin{figure*}
    \centering
    \includegraphics[width=1.0\linewidth]{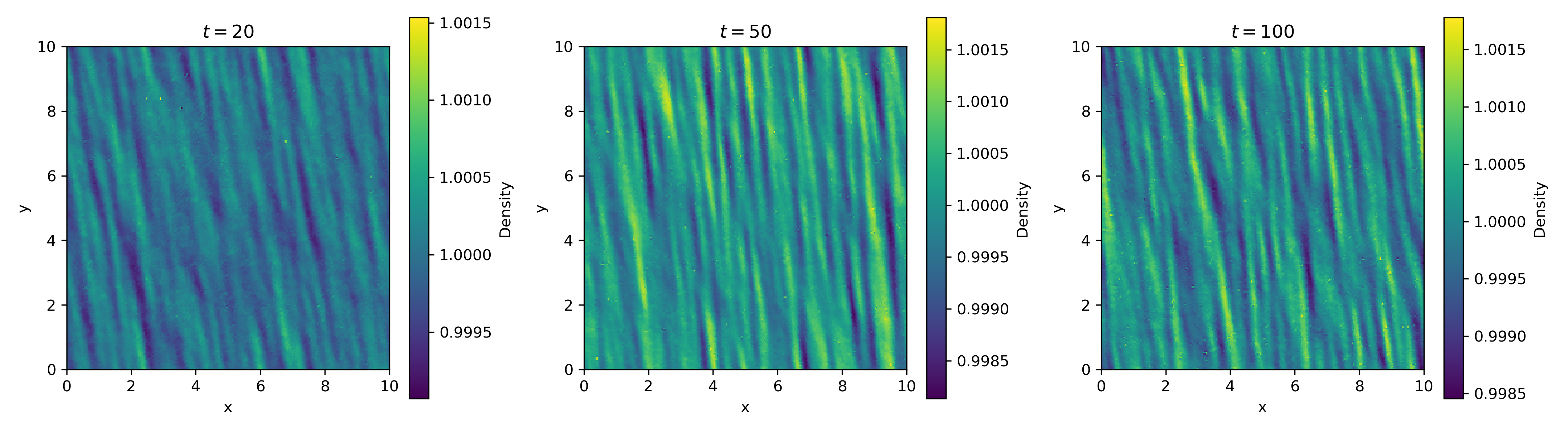}
    \caption{The density field at different times in a two-dimensional calculation that simulates the ground state of the shearing box with an initially constant density, $\rho = 1$, using the old default version of the {\small AREPO} code, as described in Section~\ref{sec:OverViewArepo}. The simulation seeds small `grid-noise' at the level of $10^{-3}$ which is stretched by the shearing motion into vertically oriented bands. This effect has a purely numerical origin and cannot be eliminated by resorting to finer timestepping.}
    \label{fig:groundStateDensityOld}
\end{figure*}

For testing this, we set up a two-dimensional box of size $L_x = L_y = 10$, density $\rho=1$, and an isothermal equation of state with sound speed $c_s = 1$ and $\Omega_0 = 1$. The initial mesh is created by two nested Cartesian grids with $200^2$ mesh-generating points, effectively creating a hexagonal mesh. We add to these positions of the mesh-generating points some random noise with a maximum of 5\% of the lattice constant of the grid, and set the velocity of the cells to the value of the ground state (\ref{eq:groundState}) at the centers of mass of the resulting mesh cells. In \cref{fig:groundStateDensityOld}, we show the time evolution of the computed density field. We see that the code is not able to stably keep the imposed ground state to machine precision, rather a time-dependent stripe pattern forms, apparently due to local noise peaks that are subsequently sheared away by the fluid motion.

Since there are no spatial scales in the problem that need to be resolved in the ground state, the only sources of error are of numerical origin. If they are related to the temporal discretization of the updates of the fluid state,  the error in the density field should decrease with the size of the time step. However, if we rerun the simulation with different Courant coefficients (see also eqn.~\ref{eq:timeStepSize}) and compare the $L_1$ errors of the resulting density fields, we find that the error does not decrease with smaller time steps sizes, i.e.~it does not arise from truncation errors of the temporal discretization and will still be present even for arbitrarily small timesteps.

This raises the suspicion that the errors are directly related to the unstructured cell geometries used in our calculation. To test for this possibility, we run our code with a static grid, although this requires a special trick to deal with the shearing box boundary conditions because the approach used for this purpose in Eulerian codes is not implemented in our moving-mesh approach. However, we can still compute test simulations of the shearing box with a static mesh by defining buffer zones on both sides of the simulation domain close to the boundaries in the $x$-direction. In every time step, we then simply override the fluid quantities in these regions and reset them to the analytically expected values. In this way, we can effectively restrict the live calculation of the code to the regions in between, allowing us to examine the sources of errors there without having to actually solve the boundary conditions problem for stationary meshes in a clean way.

Using this approach, we find that both a stationary Cartesian mesh, as well as a stationary hexagonal mesh (formed from two nested grids of mesh-generating points) are able to keep the ground state to machine precision after all. Furthermore, this ground state is also maintained with this precision if we let the mesh move but {\em prescribe} the motion such that a symmetry of the cells is always maintained (for example, in practice we start from a Cartesian mesh and set the velocities of the mesh generating points to the initial fluid velocity, and then keep them constant). In contrast, if we use asymmetric cell geometries and a prescribed motion, this accuracy immediately breaks down.  This leads to the conclusion that the symmetry of the Cartesian/hexagonal grid is a key factor in allowing the code to avoid the creation of local deviations from the ground state. In the next subsection, we will see that this behaviour is ultimately not surprising, because the symmetric cell geometries benefit from a fortuitous cancellation of flux area integration errors on opposite sides of mesh cells. Additional effort beyond the standard integration scheme is needed to achieve the same final  accuracy also for general mesh geometries.

\subsection{Higher order flux integration}

A clear potential source for inaccuracies is the approximation (\ref{eq:approxIntegralMid}) for the area integral of the flux over a cell interface. This approximation is only exact if the flux varies linearly across the interface. However, this condition is, for example, not fulfilled if there is a finite velocity gradient, since in this case, the momentum flux becomes a quadratic function in space, see eqn.~(\ref{eq:idealMHDFlux}). If a linear gradient in density and velocity exists, the maximum polynomial order of the flux function occurs for the energy flux and is forth order in space. This ultimately means that an accurate calculation of the surface integral $I_{ij}$ that is free of truncation errors requires a method that can integrate polynomials of order up to 4 exactly (or order 3 in case of an isothermal EOS, or if density gradients can be neglected).

The approximation (\ref{eq:approxIntegralMid}) can be viewed as the simplest low-order case of the general class of Gauss-Legendre integration rules, that can be used in any dimension. In two-dimensional simulations, the integral $I_{ij}$ is a line integral and can be transformed into the form:
\begin{equation}
I = \int_{-1}^1 f(x)\, {\rm d}x \approx \sum_{i=1}^n f(x_i)\, w_i,
\label{eq:1dGaussLegendre}
\end{equation}
where $f(x)$ represents the flux function in normalized coordinates with $x\in [-1,1]$ marking the position on the interface. The integral can be approximated by a weighted sum of values of $f$ obtained at different evaluations points $x_i$, the so-called Gauss points. For an appropriate choice of $x_i$ and $w_i$, the approximation can be made exact for polynomials up to a certain order $m$, but achieving higher-order $m$ also requires more function evaluations, i.e.~a higher $n$. In Appendix \ref{subsec:gauss2Dims} we give the exact values for $x_i$ and $w_i$ as implemented in our code, but for all test cases in this paper, two points, $n=2$, exact up to third order polynomials, are sufficient. For every $x_i$ we require an additional reconstruction of the states at the interface as well as a call to the Riemann solver. Since those costs are typically significantly smaller than the mesh construction  in two dimensions, we adopt this method as the (new) default one in the {\small AREPO} code.

In three-dimensional simulations, the interfaces of cells are polygons. Although methods exist that try to find the minimum amount of evaluation points for a specified target accuracy \citep[e.g.][]{mousavi2010generalized}, there exist no general equations for polygons with more than four edges that employ the minimum number of evaluation points needed for a targeted exact polynomial accuracy. We, therefore, split the interface into triangles over which we can integrate the flux accurately using standard Gauss-Legendre integration. The triangles themselves can be easily obtained during the construction of the Voronoi mesh from the Delaunay triangulation. For a third-order accurate method, we have to use 4 points per triangle, which means we require on average more than 15 flux evaluations per surface integral to realize this accuracy for arbitrary cell geometries. While this sounds like a significant cost increase, using simulations of the background shear flow (\ref{eq:groundState}) we found an overhead of only around 110\% for a naive implementation of this method. However, we were able to reduce this overhead to around 30\% by performing the time extrapolation only once per face and not for every evaluation point separately, and by an explicit vectorization of the flux calculation with AVX-2 instructions available on modern CPUs. In Appendix~\ref{subsec:gauss3Dims} we discuss further details of the application of the Gauss-Legendre integration in three dimensional Voronoi meshes.

\subsection{Less strict slope limiter}

In regions with a smooth flow, the linear reconstruction of equation~(\ref{eq:reconstructionGeneral}) leads to a second-order accurate method in space. But in regions with physical discontinuities, the standard reconstruction scheme can induce under- or overshoots, and thus oscillatory behaviour, unless a non-linear slope limiter is applied to reduce the local gradients such that no new extrema are introduced into the solution. As  this reduces the order of the method to first-order accuracy close to discontinuities, it is important to choose a limiting scheme that only reduces the gradient where really required, and by the least amount needed to  accurately capture shocks and contact discontinuities, and to prevent numerical instabilities.

Thus far, {\small AREPO} by default uses the scheme from  \cite{springel2010pur}: The original estimate of the gradient of a quantity $\phi$ is replaced by 
\begin{equation}
\left<{\vec{\nabla}}\phi\right>_i^{'} = \alpha_i
\left<\vec{\nabla}\phi\right>_i ,
\end{equation}
 where $0 \leq \alpha_i \leq 1$ is the value of the limiter of the given cell. It is computed as
 \begin{equation}
 \alpha_i = \min_j(1, \psi_{ij}) ,
 \end{equation}
where $\psi_{ij}$ is here defined for every interface of the cell $i$ as
\begin{equation}
\psi_{ij} = \left\{ 
\begin{array}{ccc}
(\phi_i^{\rm max} -  \phi_i)/ \Delta\phi_{ij}  & {\rm for} & \Delta\phi_{ij}>0\\
(\phi_i^{\rm min} -  \phi_i)/ \Delta\phi_{ij}  & {\rm for} & \Delta\phi_{ij}<0\\
1 & {\rm for} & \Delta\phi_{ij}=0\\
\end{array}
\right.
\end{equation}
and $\Delta\phi_{ij}=\left<{\nabla}\phi\right>_i\cdot(\bm {f}_{ij} - \bm {s}_i)$ is the estimated change between the centroid $\bm {f}_{ij}$ of the face and the centre of cell $i$, while $\phi_i^{\rm max} = \max(\phi_j)$ and $\phi_i^{\rm min} = \min(\phi_j)$ are the maximum and minimum values occurring for $\phi$ among all neighbouring cells of cell $i$, including $i$ itself.

However, this scheme can lead to a too restrictive limiting of the gradient of the background shear flow (\ref{eq:groundState}) for an unstructured mesh. This can happen for special cell geometries where the face connecting two cells is displaced quite far from the line connecting the corresponding mesh generating points. As a result, one can end up with $\alpha_i < 1$ even if there is no discontinuity in the flow. But this in turn increases the local truncation error of the solution, introducing unwanted `grid-noise' into the solution. To prevent this, we replace $\bm f_{ij}$ in the above formulation of the slope limiter by the mid point $\frac{1}{2} (\bm s_i + \bm s_j)$ of the center of mass of the cells $i$ and $j$. We found that this formulation prevents post-shock oscillations close to discontinuities in all our test simulations just as well as the original formulation, while it safely avoids being triggered in situations with a uniform linear gradient in combination with a highly irregular Voronoi mesh.

\begin{table}
\begin{center}
\begin{tabular}{ |c|c|c|c| } 
 \hline
 simulation & initial & improvements & order time \\ 
    name &  grid & enabled? & integration \\ 
  \hline
 CO2 & cart. & no & 2 \\ 
 PO2 & polar & no & 2 \\ 
 CI2 & cart. & yes & 2 \\ 
 PI2 & polar & yes & 2 \\ 
  CI3 & cart. & yes & 3 \\ 
 PI3 & polar & yes & 3 \\
  \hline
\end{tabular}
\end{center}
\caption{Overview of the different code configurations used for the isentropic Yee vortex test. The simulations differ in the topology of the initial grid layout (with is either Cartesian or polar), whether or not the improvements (but without higher-order time integration) presented in this paper are enabled, and finally whether a second or third order time-integration is used.}
 \label{tab:yeeOverview}
\end{table}

\begin{figure}
    \centering
    \includegraphics[width=1.0\linewidth]{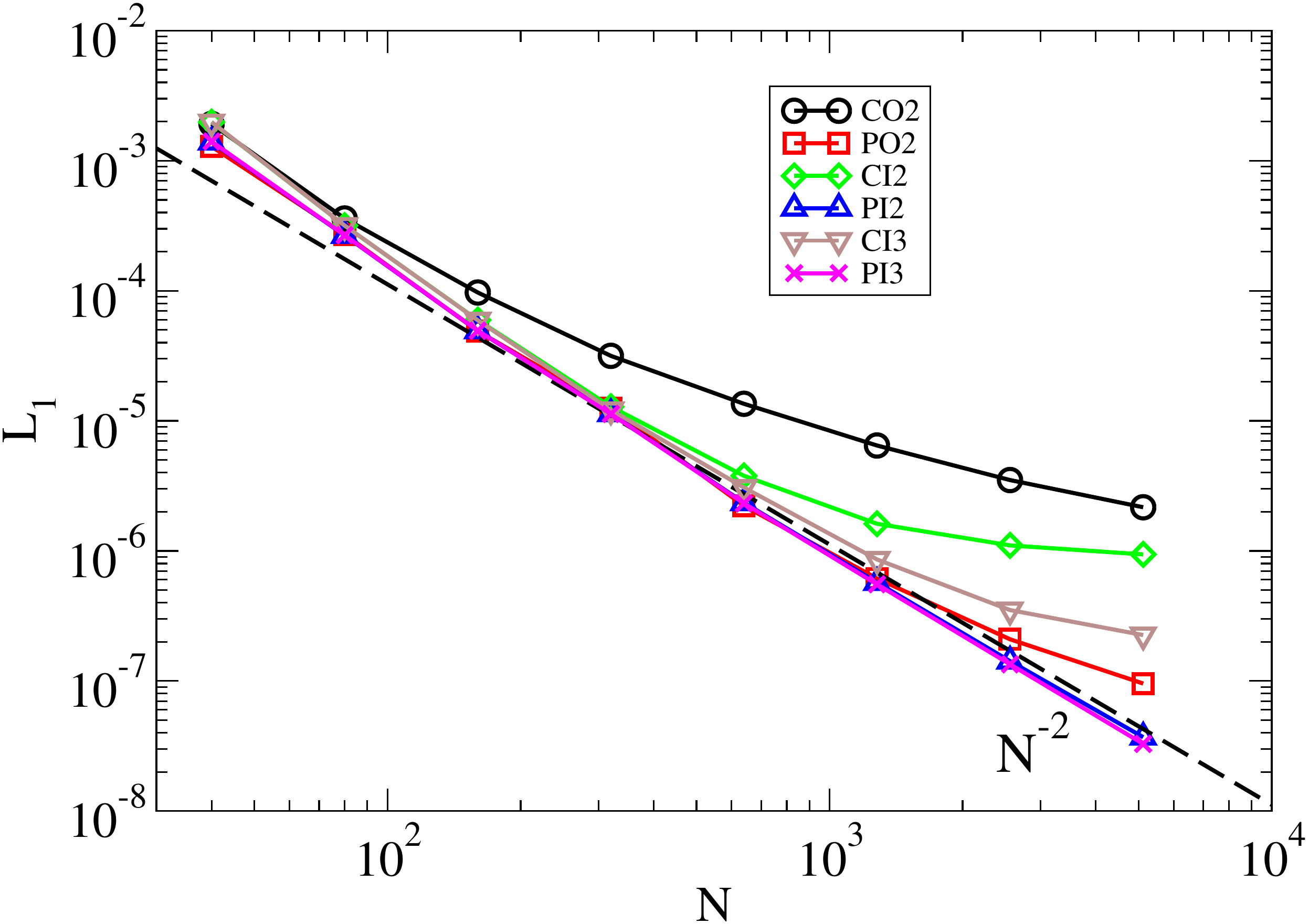}
    \caption{$L_1$ norm of the error of the density field for the Yee vortex at simulation time $t = 10$ when evolved with different hydrodynamical schemes, as defined in Table~\ref{tab:yeeOverview}. Our improved version of {\small AREPO} robustly obtains second-order convergence independent of the initial grid geometry, whereas the old default version achieves this only for polar grids over a limited range of resolutions. }
    \label{fig:YeeConvergence}
\end{figure}

\subsection{Improved temporal and spatial extrapolation}

Equation (\ref{eq:reconstructionGeneral}) does not specify in which frame we perform the spatial reconstruction and temporal extrapolation. The simplest choice would be the lab frame, in this case, $\bm s = \bm s_{\rm old}$ is the centre of mass at the time we calculate the gradient and not the current centre of mass.  A more natural choice would be the moving frame of the gas. In this case we use $\bm s = \bm s_{\rm old} + \bm v \Delta t$ as the origin of the spatial reconstruction and set $\bm v = 0$ in equation (\ref{eq:linearTimeExtrapolation}). We realized through our sensitive tests that the current version of {\small AREPO} implemented the reconstruction and extrapolation in the moving frame in a slightly inconsistent way. It transformed into the moving frame of the interface and performed the reconstruction from the current centre of mass. 

Since the velocity of the cell is close to the velocity of the gas and the velocity of the frame is close to the velocity of the adjacent cells, this generally only leads to small errors.  These are however especially strong (and superfluous) in cases where  strong mesh regularisation motions are applied, and then can be observed in simulations of the ground state of the shearing box as spurious noise. This noise is now eliminated in our improved version of the code.

\subsection{Higher-order Runge-Kutta time integration}
\label{subsec:higherOrderRK}

The approximation~(\ref{eq:TimeIntegrationGeneral}) for the time integration is in general second order accurate, which means that its error $E_{t,s}$ for a single time step scales a  $E_{t,s} \propto \Delta t^{3}$ with timestep size, and the accumulated error for the integration over a fixed time interval scales as $E_{t,i} \propto \Delta t^2$. In many simulations, the total error is typically dominated by the error $E_{x}$ of the spatial discretization, which should be proportional to the square of the local cell size $\Delta x$ in smooth flows. 

This is because the size of the time step and the local spatial resolution are connected by the stability criterion:
\begin{equation}
    \Delta t = C_{\rm CFL} \frac{\Delta x}{v_{\rm signal}},
     \label{eq:timeStepSize}
\end{equation}
where $v_{\rm signal}$ is the local signal speed and $C_{\rm CFL}$ the Courant factor as a free parameter. This implies that an increase in the spatial resolution reduces also the error of the time integration by a comparable factor, and the whole scheme converges with second order, typically also keeping the time integration error subdominant.

\begin{figure*}
    \centering
    \includegraphics[width=0.8\linewidth]{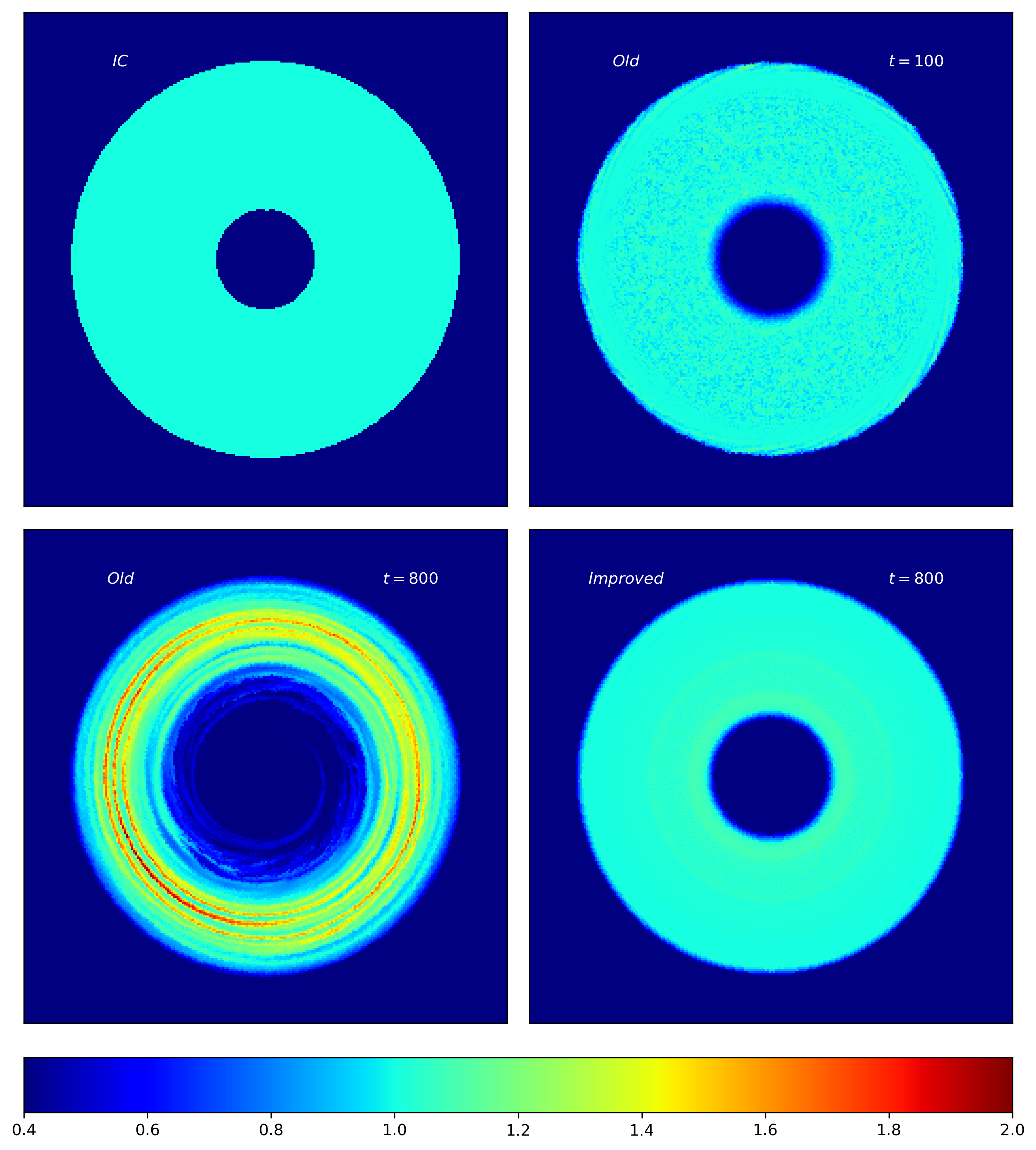}
    \caption{The density distribution of a cold Keplerian disk evolved with different simulation methods and at different times, as labelled. The method marked `old' corresponds to the default version of {\small AREPO} without the modifications proposed here, while the run marked 'improved' contains all modifications except for the third order time integration. We here use a polar grid for the initial conditions.}
    \label{fig:Kepler800}
\end{figure*}

\begin{figure*}
    \centering
    \includegraphics[width=0.7\linewidth]{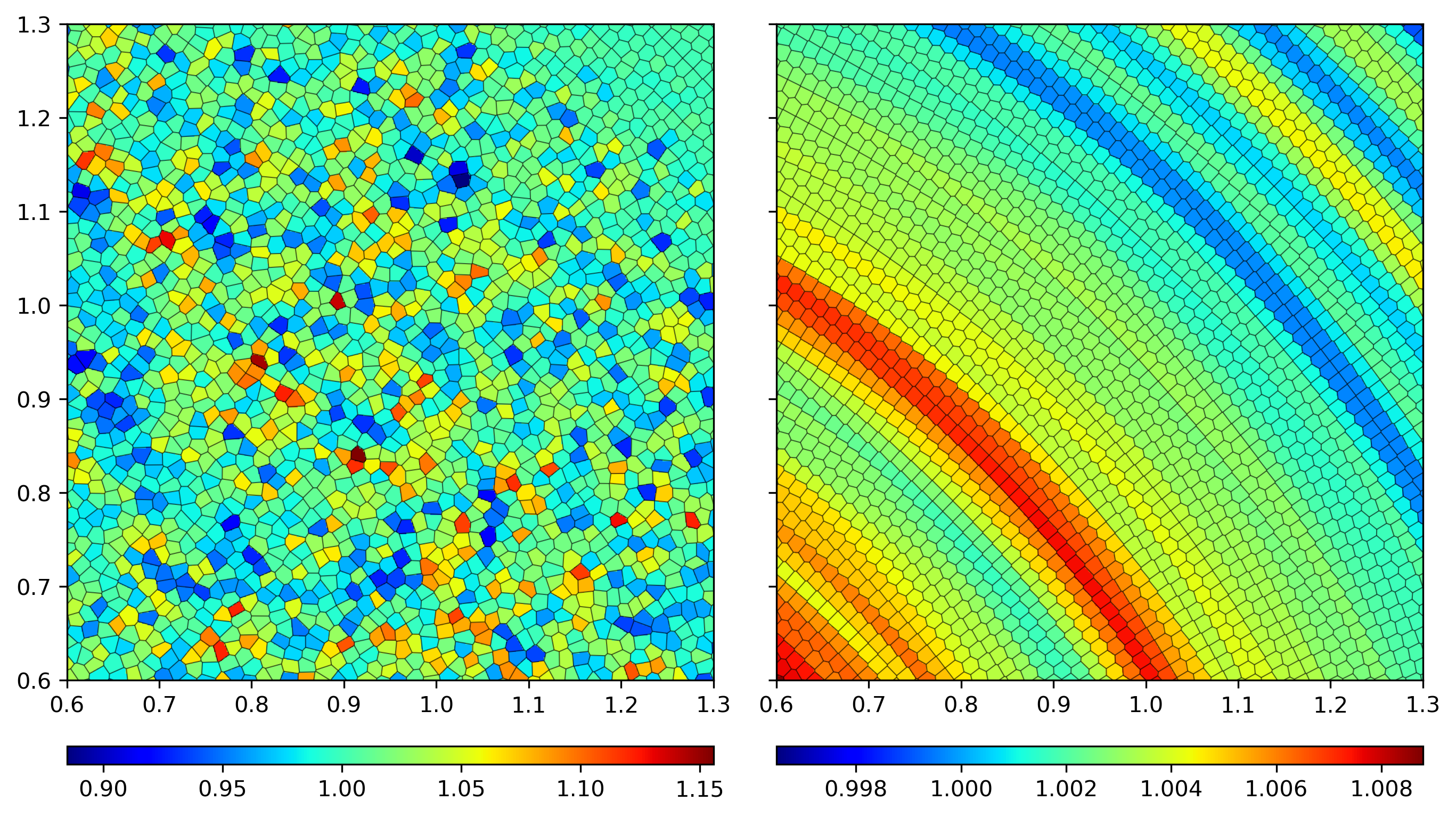}
    \caption{A zoom into the density field for a Keplerian disk simulation at time $t=100$. The left hand side shows the disk evolved without the code improvements proposed here while the right hand side includes the modifications.  Both simulations were initialized with a polar grid of mesh-generating points.  The right simulation has much smaller, spherical, residual density perturbations (note the different stretch of the color tables).}
    \label{fig:KeplerZoom}
\end{figure*}

\begin{figure*}
    \centering
    \includegraphics[width=0.8\linewidth]{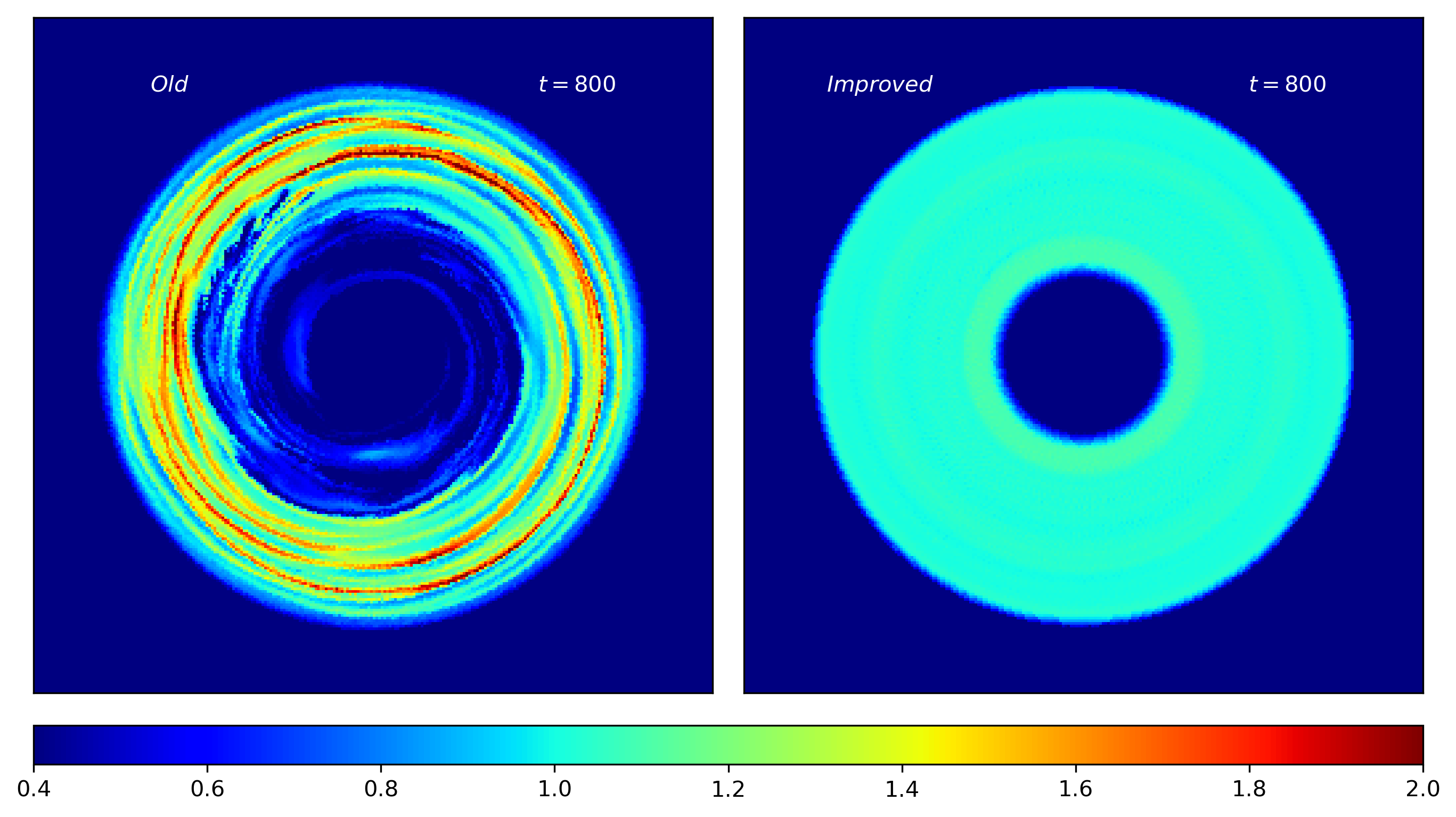}
    \caption{The same as the lower row of panels in \cref{fig:Kepler800} but for a simulation with an initially Cartesian grid and not a polar grid. Even in this case, our improved simulation code exhibits a drastically reduced noise level in this setup, and thus a much better long-term stability of the cold, rotationally supported disk.}
    \label{fig:Kepler800Cartesian}
\end{figure*}

Since there are no spatial scales to resolve in the ground state of the shearing box (see Section~\ref{subsec:groundState}), the error  in this case is however dominated by the error in the time integration, and can be observed as noise in the unstructured grid case. To verify this and reduce this noise if desired, we also implemented a formally third order accurate Runge Kutta method for the integration of the ideal MHD equations:
\begin{equation}
    \bm Q_i^{n+1} = \bm Q_i^n +\frac{\Delta t}{6} \left[\bm I(t_0) + 4 \bm I\left(t_0+ \frac{\Delta t}{2}\right)+  \bm I(t_0+ \Delta t)\right].
    \label{eq:timeIntegration3rdOrder}
\end{equation}
This adds an extra mid-timestep evaluation of the integral $\bm I$, for which we use again the linear time extrapolation (\ref{eq:linearTimeExtrapolation}). We note that other parts of the code, for example the source terms, are still only integrated with the second-order accurate method (\ref{eq:TimeIntegrationGeneral}), which means that if those terms are dominating the error, the error will still only decrease with second order in time.

\subsection{Testing the accuracy improvements}

Although all of the modifications of {\small AREPO} presented in this section were developed specifically to reduce the noise of the background flow in a shearing box, they are also useful for other applications of the code. In this subsection, we will briefly examine some representative examples to highlight this welcome effect.

\subsubsection{Yee vortex}
\label{subsubsec:YeeVortex}

The isentropic Yee vortex test consists of a rotating, quasi-stationary flow with a differentially shearing velocity profile. Since the flow is smooth and free of discontinuities, it can be used to verify second-order convergence for a non-trivial two-dimensional flow. The test was used in \cite{pakmor2016improving} to show that the very first version of {\small AREPO} \citep{springel2010pur} did not manage to guarantee full second-order convergence for this problem due to a comparatively noisy gradient estimation scheme and inaccuracies in time integration. \cite{pakmor2016improving} introduced improvements to both points that have since been used by default in the code. As shown in \cite{pakmor2016improving}, using this updated code the Yee vortex then converges with second order, but this still requires a specially prepared polar grid whereas a start from a cubical grid that is ignorant of the spherical symmetry of the mesh motion still spoils the high convergence rate. 

We here return to examining this issue and rerun the Yee vortex with and without the improvements presented in the previous subsections. For definiteness, we use both an initially polar grid as well as a Cartesian grid. Also, we run the test simulations with the standard second order accurate time integration (\ref{eq:TimeIntegrationGeneral}) as well as with the improved third order accurate one (\ref{eq:timeIntegration3rdOrder}). The details of the initial conditions can be found in Appendix~\ref{subsec:setupYee}, and in Table~\ref{tab:yeeOverview} we give a schematic overview of the different test runs.

To measure the convergence rate we define the $L_1$ error as
\begin{equation}
L_1=    \frac{1}{V} \sum_{i=1}^{N_{\rm cell}} |f_i|,
\end{equation}
where $|f_i|$ is the deviation of the density of a cell $i$ from the theoretical value at the coordinate of its center of mass. In \cref{fig:YeeConvergence}, we show the error norm as a function of spatial resolution for our different simulation schemes. 

We confirm that the previous version of the {\small AREPO} code was indeed only able to achieve second order convergence for this problem if an optimized polar grid (PO2) is used for the initial conditions, while a Cartesian grid (CO2) shows substantially higher errors. Interestingly, for very high resolution ($5120^2$~cells) we also find for the polar grid (PO2) a slight decrease in the order of the method. This effect vanishes if we use our improved formulation for the polar grid (PI2 and PI3) because this eliminates spurious activations of the slope limiter. 

If we start with a Cartesian grid instead, our new formulation (CI2) achieves second-order convergence 
for low to moderate resolutions, but this becomes worse again for very high resolution. This degradation at very high resolution can be avoided by using our new  improved time integration accuracy (CI3), suggesting that for low to moderate resolution the total error is dominated by errors of the spatial discretization, while for very high resolution the total error becomes dominated by  the temporal discretization. The reason for this is presumably related to the local mesh regularization that we apply, and a faster change of the local topology of the grid when the resolution is higher.

\subsubsection{Keplerian disc}

A challenging problem for many hydrodynamical methods is the evolution of a pressure-less gas disc orbiting around a central object on a Keplerian orbit \citep{cullen2010inviscid, hopkins2015new, pakmor2016improving}. In particular, \cite{hopkins2015new} reported that many state-of-the-art SPH and mesh codes show severe problems in this test. However, using their new MFM method they were able to evolve the disk for more than 250 inner orbits ($t=600$). Although the disk showed noise in the density evolution of the order unity, the disk seemed to stay stable.

\cite{pakmor2016improving} used the same test to show that {\small AREPO} with his improved gradient estimates  was also able to stably evolve the disk until $t=600$.  But at this time the disk started to break up at its inner boundary, and during the whole evolution, some noise in the density profile of the disk could be observed.

We rerun this test with the improved methods presented in the previous subsections (except for the higher-order time integration), with the exact same initial conditions as \cite{pakmor2016improving} (for details see Appendix~\ref{subsec:setupKepler}) and at a resolution of 320x320 cells. As we show in \cref{fig:Kepler800}, we reproduce their result that  the original version of {\small AREPO} is able to evolve the disk for several hundred code units, and that at around $t=600$ the inner boundary of the disk starts to break up, which only becomes worse with time. 
However, the improved integration method we proposed earlier  allows us to evolve the disk for a much longer time. In fact, even at $t=800$ the disk remains stable and only shows a small increase in density at the inner edge of the disk due to a weak influence of the inner boundary condition. In addition, the new scheme at the same time significantly reduces the noise in the density field throughout the disk.

In \cref{fig:KeplerZoom} we show a zoom into the density field,  including the geometry of the Voronoi mesh at $t= 100$.  While in the simulation with the old scheme the initial polar grid is destroyed and the local noise has a relative size of 10\%, the initial grid stays stable in the simulation with the improved scheme. The maximum deviation from the theoretical density field is only around 1\% (observe that the color scale is stretched accordingly) and those deviations do not take the form of local `grid noise', but rather are ring-like structures that are coherent for the whole disk.

To further analyze the sensitivity of this simulation problem to the initial grid layout, we rerun  simulations with an initial Cartesian grid of size $320 \times 320$ cells, once with the old code and once with our improved method. In  \cref{fig:Kepler800Cartesian} we show the density distribution at $t=800$. In the simulation without the improvements proposed in this paper, the disk starts to break up earlier compared to the simulation started with an initial polar grid. In contrast to this, the simulation with the improvements looks very similar to the simulation with an initially polar grid, and thus is insensitive to the initial grid geometry, as desired.

\section{Test problems for the shearing box}
\label{sec:testProblems}

In this section, we analyze the performance of our implementation of the shearing box approximation in several test cases. We consider many of the problems used in \cite{stone2010implementation} to analyze the accuracy of the shearing box implementation  in the {\small ATHENA} code, and we employ in several cases similar initial conditions as they did. As in Section~\ref{subsubsec:YeeVortex} we define the $L_1$ error of a quantity $f$ as
\begin{equation}
L_1 =    \frac{1}{V} \sum_{i=1}^{N_{\rm cell}} |f_i|,
    \label{eq:L1DefintionAverage}
\end{equation}
where $|f_i|$ is the difference of the value of a cell $i$ from the theoretical value of $f$ at its center of mass.

Although the use of a perfect Cartesian grid sometimes leads to better results due to the additional symmetry in the mesh configuration, we deliberately avoid this by adding a random displacement of 2\% of the mean particle spacing to the positions of the  mesh-generating points in the initial conditions. The background shear flow, together with mesh regularisation motions introduced by the code, then rapidly create a fully unstructured mesh, which is more representative for the mesh configurations encountered in realistic production runs with {\small AREPO}. To initialize our test simulations, we set the fluid properties of a cell to the value assumed by the continuous fields at the coordinate of the centre of mass of the cell. If not stated otherwise, we impose a global time step for all cells. As default shear parameters we use $q = 3/2$ and $\Omega_0 = 1$, and for all isothermal simulations we use an isothermal sound speed of $c_s = 1$.

\subsection{Ground state of the shearing box}

\label{subsec:groundState}
In Section~\ref{subsec:groundStateSimulationOld} we showed that the default version of {\small AREPO} fails to accurately simulate the ground state of the the shearing box, prompting us to develop several accuracy improvements of the code. We now revisit this problem in this section, but with these improvements enabled. As a first initial step, we use an {\em unstructured static} mesh with buffer zones close to the $x$-direction boundaries so that we can ignore the shear-periodic boundary conditions. In this case, the ground state is now stable up to machine precision, implying that the fluxes are calculated exactly by the higher-order Gauss-Legendre integration we introduced.

Next, we allow the mesh to move arbitrarily, so that the geometries of the surfaces of the cells change continuously in time. This change does not have to be a polynomial function in time, which means that the second-order accurate time integration scheme of equation~(\ref{eq:TimeIntegrationGeneral}) is associated with small errors. These can be observed as small noise in the velocity and density fields that lies substantially above machine precision. The noise can be decreased by either using a smaller time step, or by using a higher-order time integration scheme, for example the third-order Runge Kutta method we described in Section~\ref{subsec:higherOrderRK}. 

To analyze the effect of the time step size and integration scheme, we use a two-dimensional box of size $L_x = L_y = 10$, and set up a Cartesian grid with $200^2$ cells and random displacements of 1\% in the coordinates of the mesh-generating points. The initial conditions are given by the ground state of the shearing box, and we run the simulation until time $t= 1000$. This leads to a dynamic unstructured mesh with a quasi stationary equilibrium between the mesh regularisation motions introduced by the code and the tendency of the local shear to degrade the local mesh regularity. We then use this ``relaxed'' unstructured mesh geometry and reinitialize the fluid variables with the analytical values of the ground state at its centre of mass to start out once more with a perfectly quiet state. Afterwards, we let the system evolve for time $t=1$ and examine how accurately the quiet shear flow of the ground state is maintained.

\begin{figure*}
    \centering
    \includegraphics[width=0.42\linewidth]{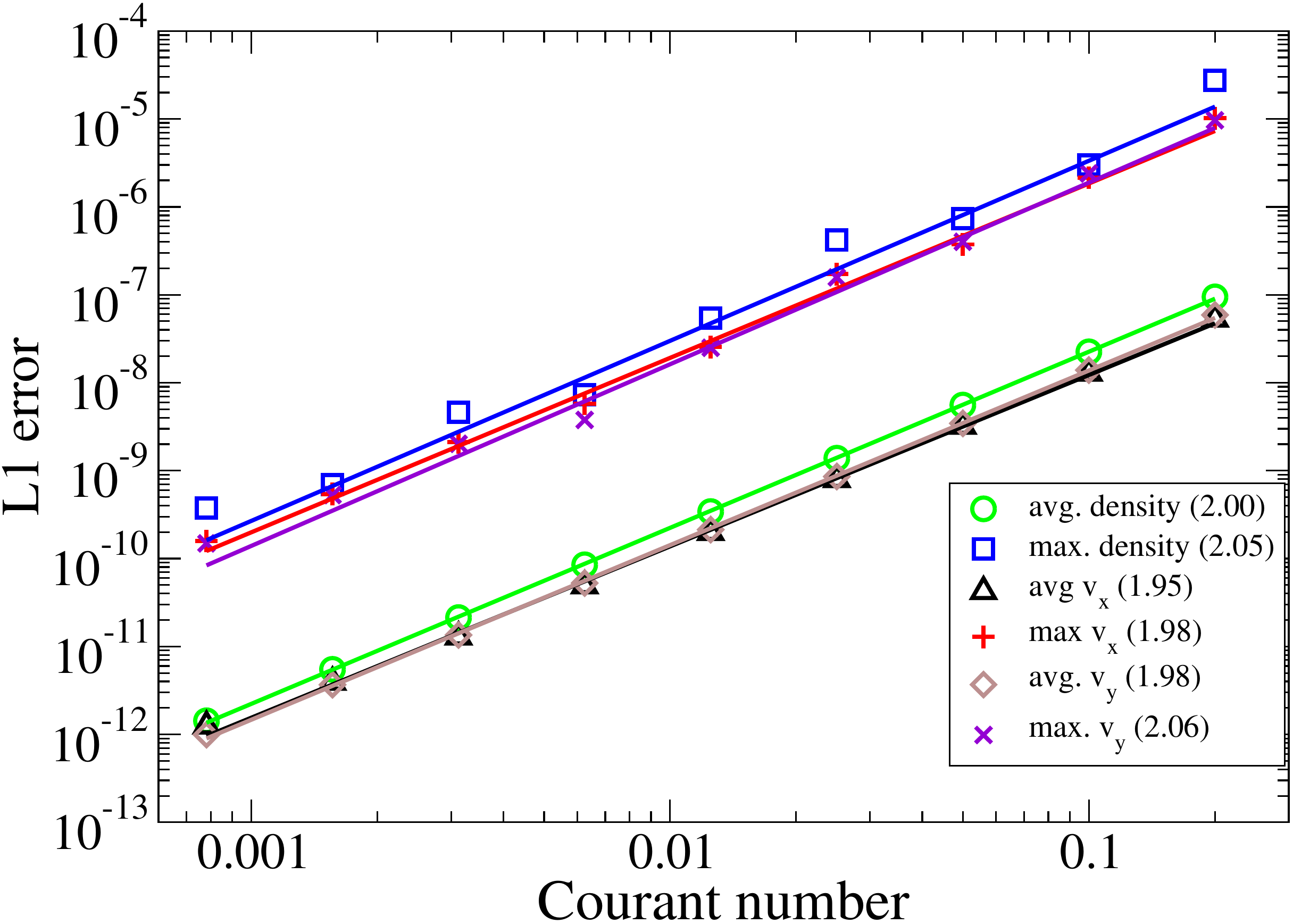}
     \includegraphics[width=0.42\linewidth]{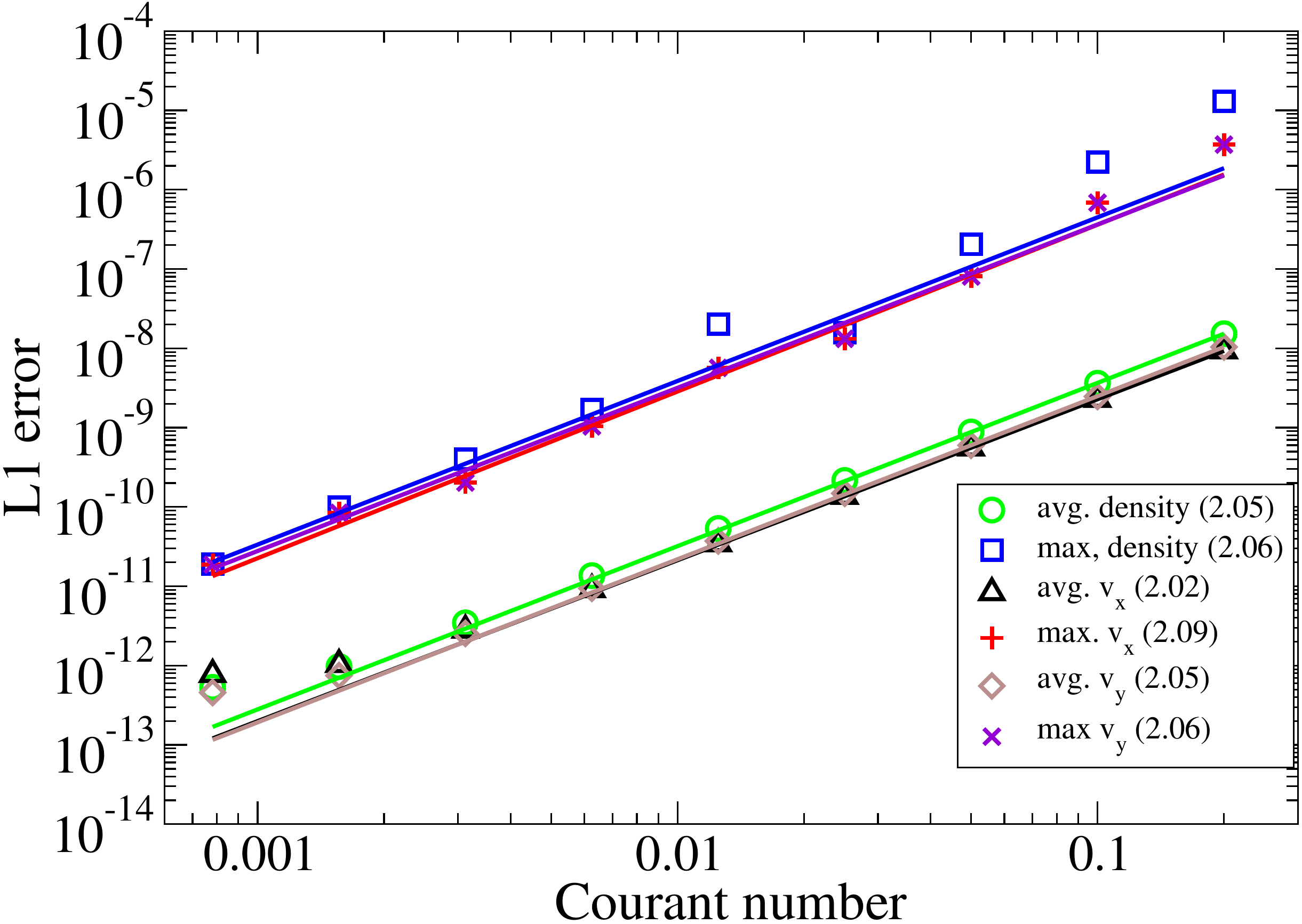}\\
         \includegraphics[width=0.42\linewidth]{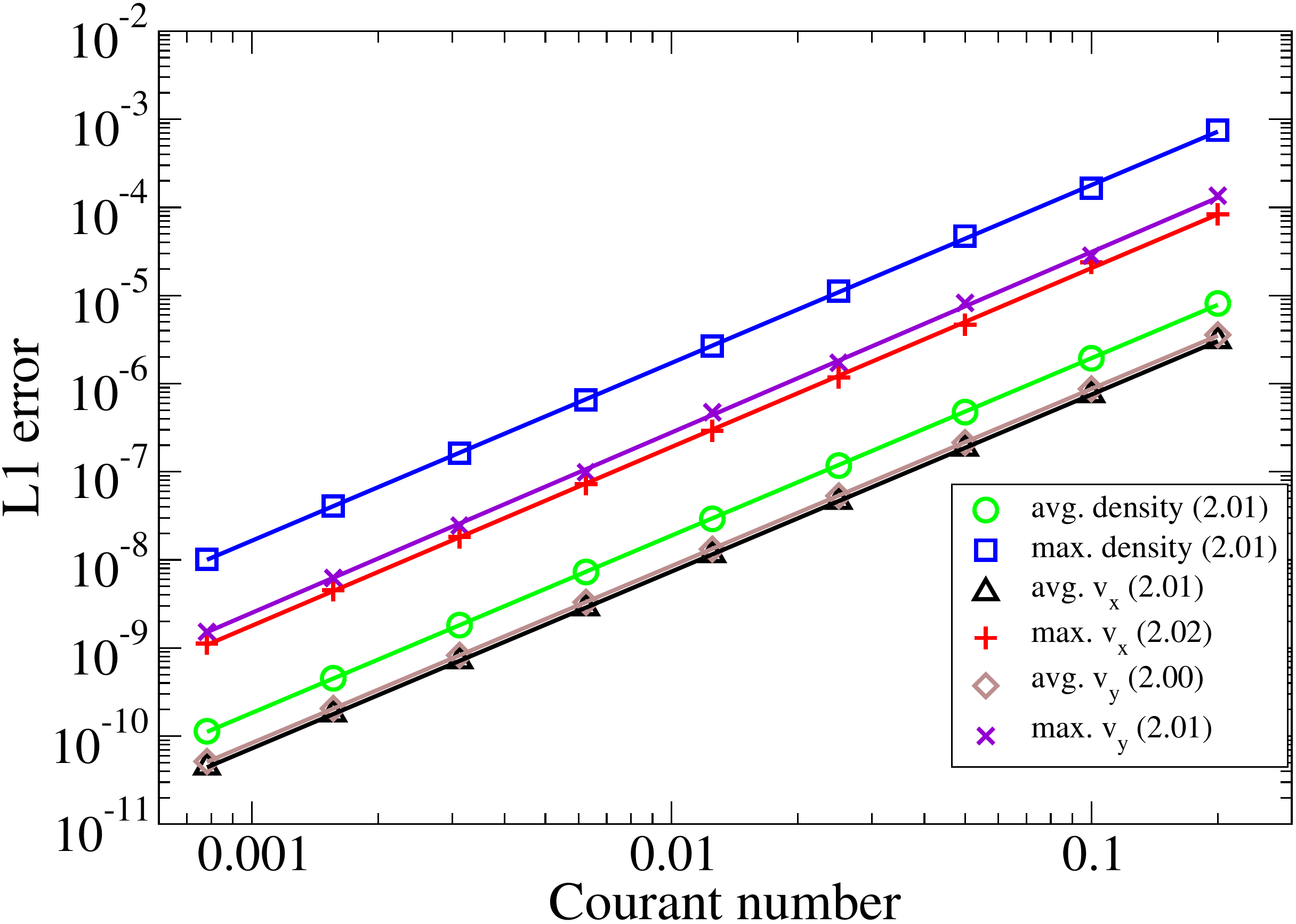}
     \includegraphics[width=0.42\linewidth]{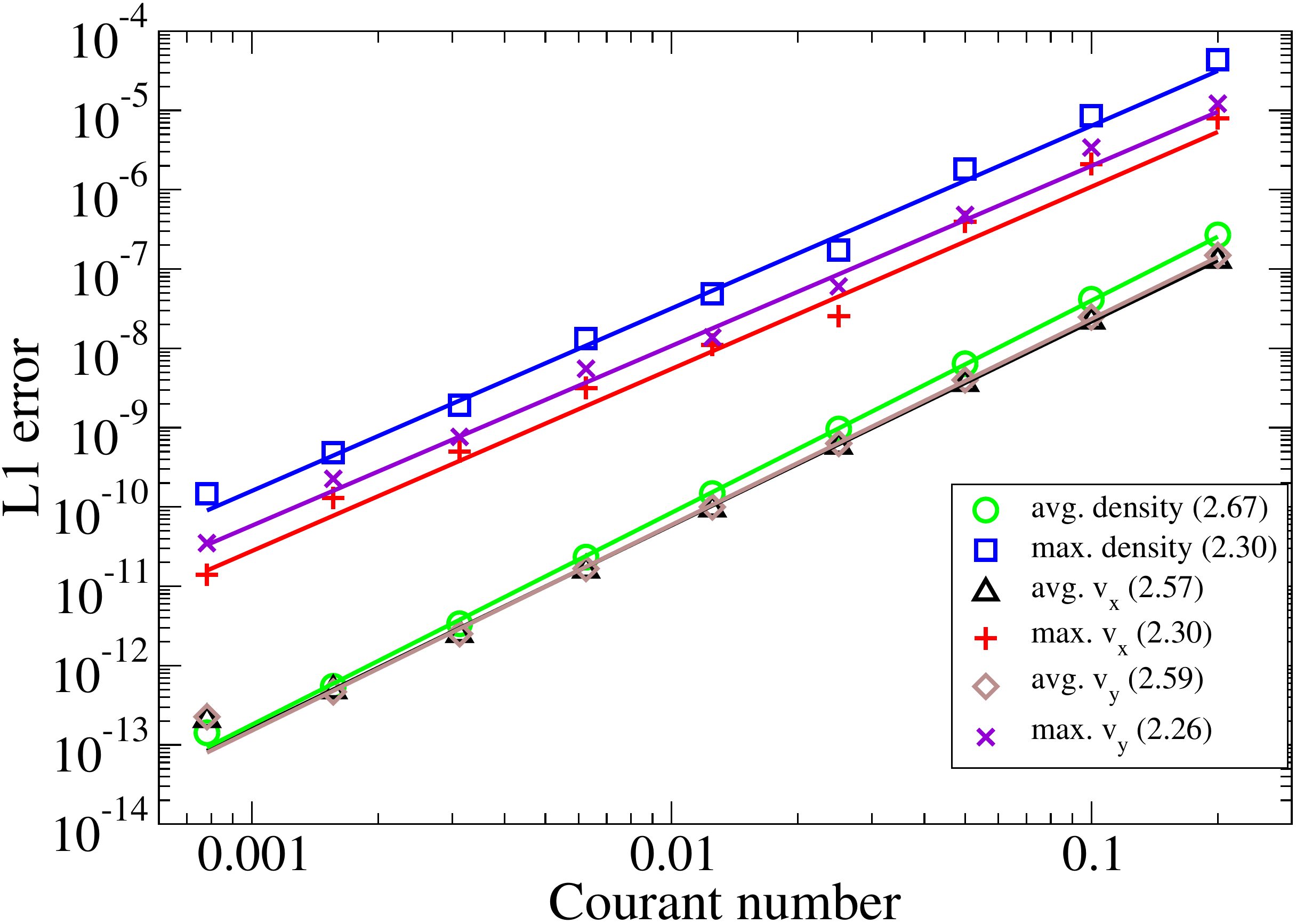}
    \caption{The $L_1$ error norm of different hydrodynamic quantities (as labelled) as a function of the time step size (here parameterized through the Courant number) at time $t=1$ for simulations of the ground state of the shearing box (for details, see Section~\ref{subsec:groundState}).  The upper row shows results in two dimensions, the lower one in three dimensions.  The panels on the left use the second order accurate time integration of eqn.~(\ref{eq:TimeIntegrationGeneral}), the ones on the right the third order accurate method of eqn.~(\ref{eq:timeIntegration3rdOrder}). In brackets we give the fitted values of the slope $A_1$, see eqn.~(\ref{eq:FitConvergenceOrder}).}
    \label{fig:groundStateL1Courant}
\end{figure*}

We repeat this last simulation for different time step sizes, imposed here by changing the Courant number. We then measure the average $L_1$ error of the fluid quantities, as well as the maximum $L_1$ error that is realized for any of the cells. In the first row of \cref{fig:groundStateL1Courant}, we show the $L_1$ errors of different hydrodynamic quantities as a function of the Courant number for the second and third-order Runge Kutta schemes. We also fit the expression
\begin{equation}
    L_1 = A_0\, C_{\rm CFL}^{A_1}
    \label{eq:FitConvergenceOrder}
\end{equation}
to the measurements. $A_1$ gives us the order of convergence of the corresponding fluid quantity. As expected for a second-order method, all errors decrease with second order for smaller time step sizes. This also shows that the error in this problem is indeed fully dominated by errors in the time integration, not the spatial integration, because the latter is eliminated by our higher order flux integration. The use of the third-order RK method reduces the amplitude of the total error ($A_0$) by about one order of magnitude, but the convergence rates ($A_1$) stay still close to 2 and do not approach the value of 3 we may have expected. To further analyze this discrepancy, we have repeated the test in three dimensions with $L_x = L_y = L_z = 10$ and $50^3$ cells.

\begin{figure*}
    \centering
    \includegraphics[width=0.76\linewidth]{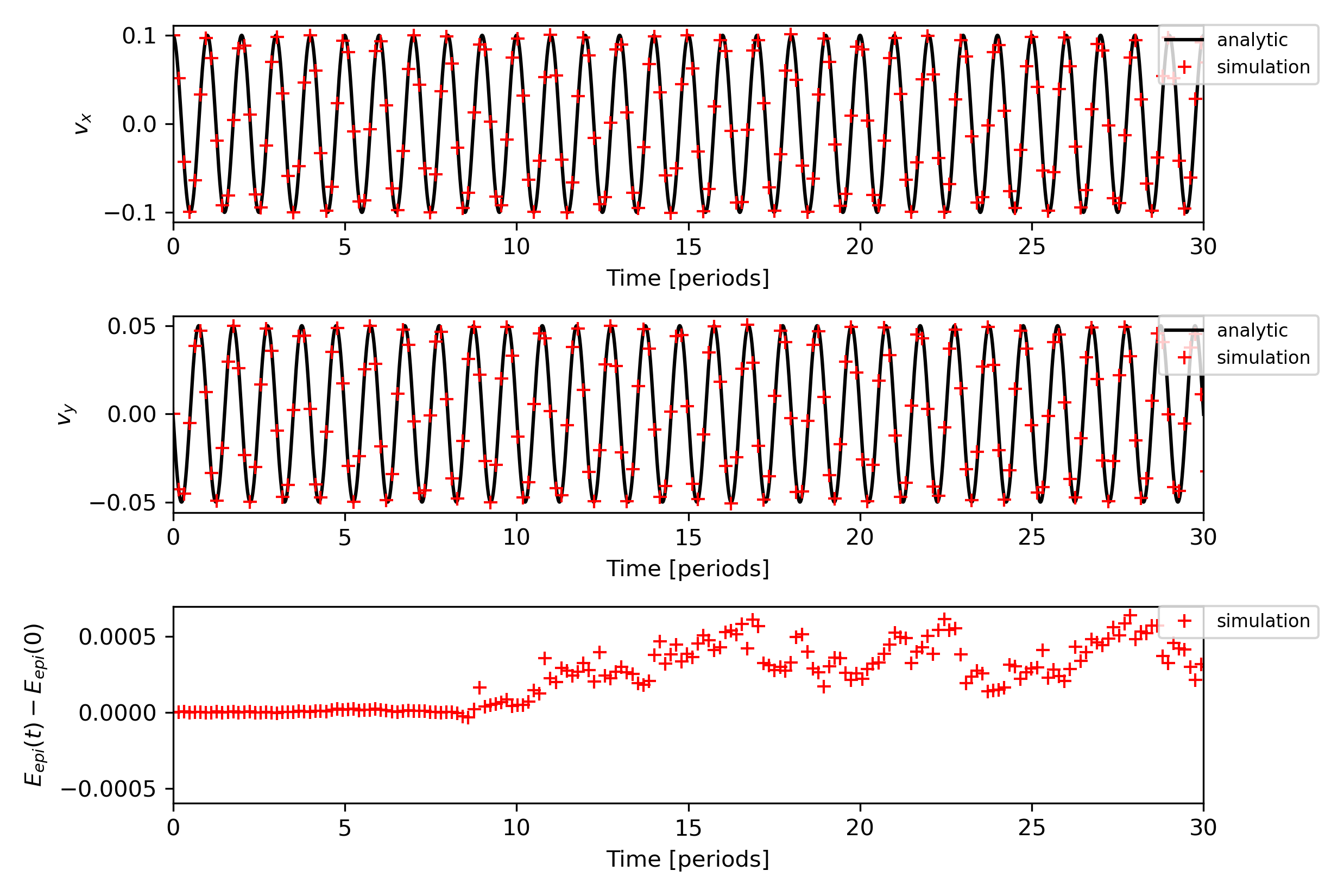}
    \caption{Epicycle oscillations in the velocities $v_x$ and $v_y$  (first and second row panels) compared between analytic solution (solid lines) and numerical  measurements (crosses) for a simulation with a resolution of $10^2$ mesh-generating points. The temporal evolution of the epicycle energy (egn. \ref{eq:epicycleOsciEnergy}) in the simulation is shown in the bottom panel.}
    \label{fig:EpicycleAmplitude10}
\end{figure*}

\begin{figure}
    \centering
    \includegraphics[width=0.9\linewidth]{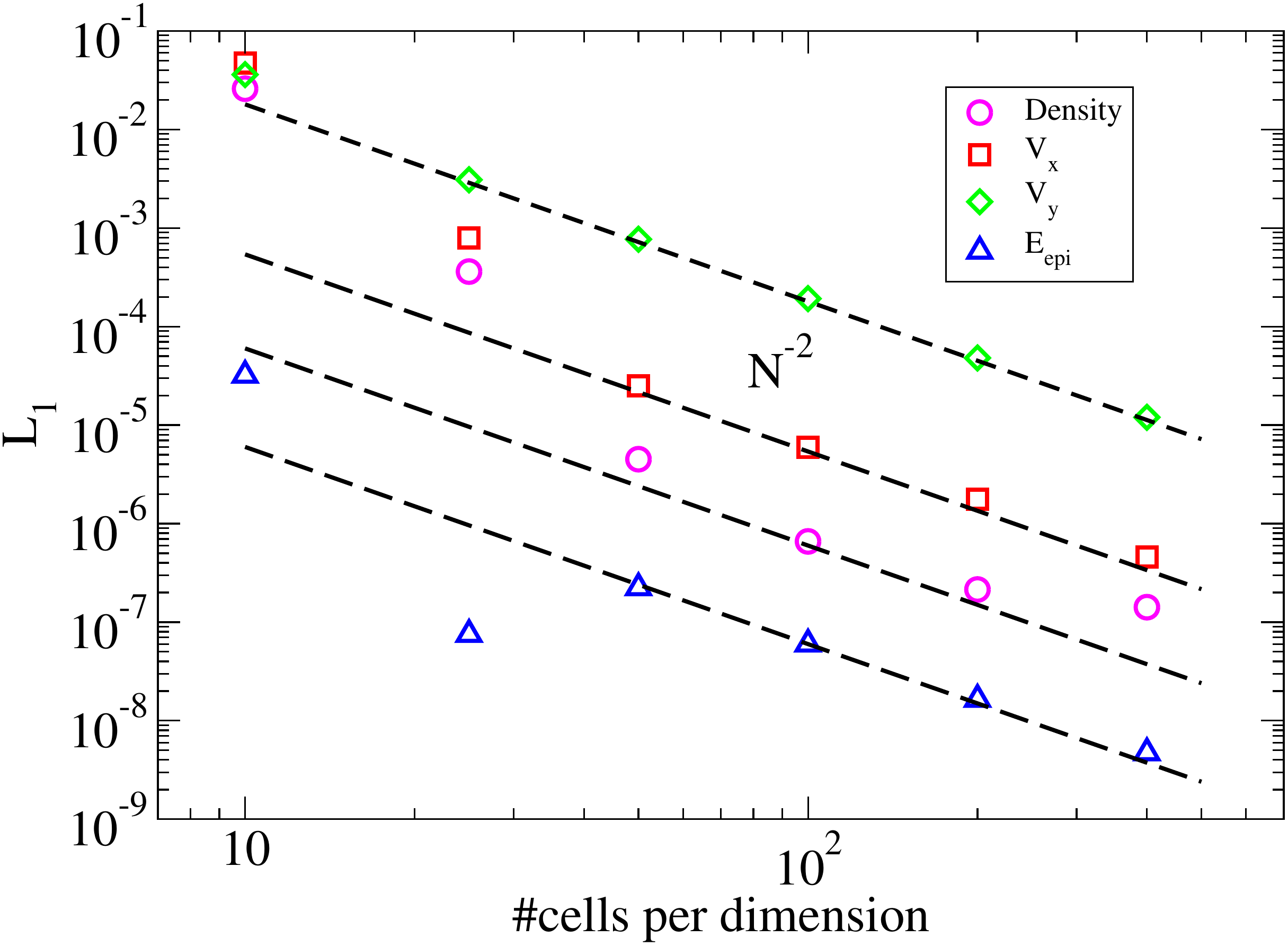}
    \caption{$L_1$ norm of the error of the density field, the velocity fields $v_x$ and $v_y$,  as well as of the epicycle energy (eqn.~\ref{eq:epicycleOsciEnergy}) for a test simulation of epicycle oscillations. The measurements were performed at time $t=666$, which corresponds to around 103 completed epicycle oscillations.}
    \label{fig:EpicycleCongergence}
\end{figure}

\begin{figure}
    \centering
    \includegraphics[width=1.0\linewidth]{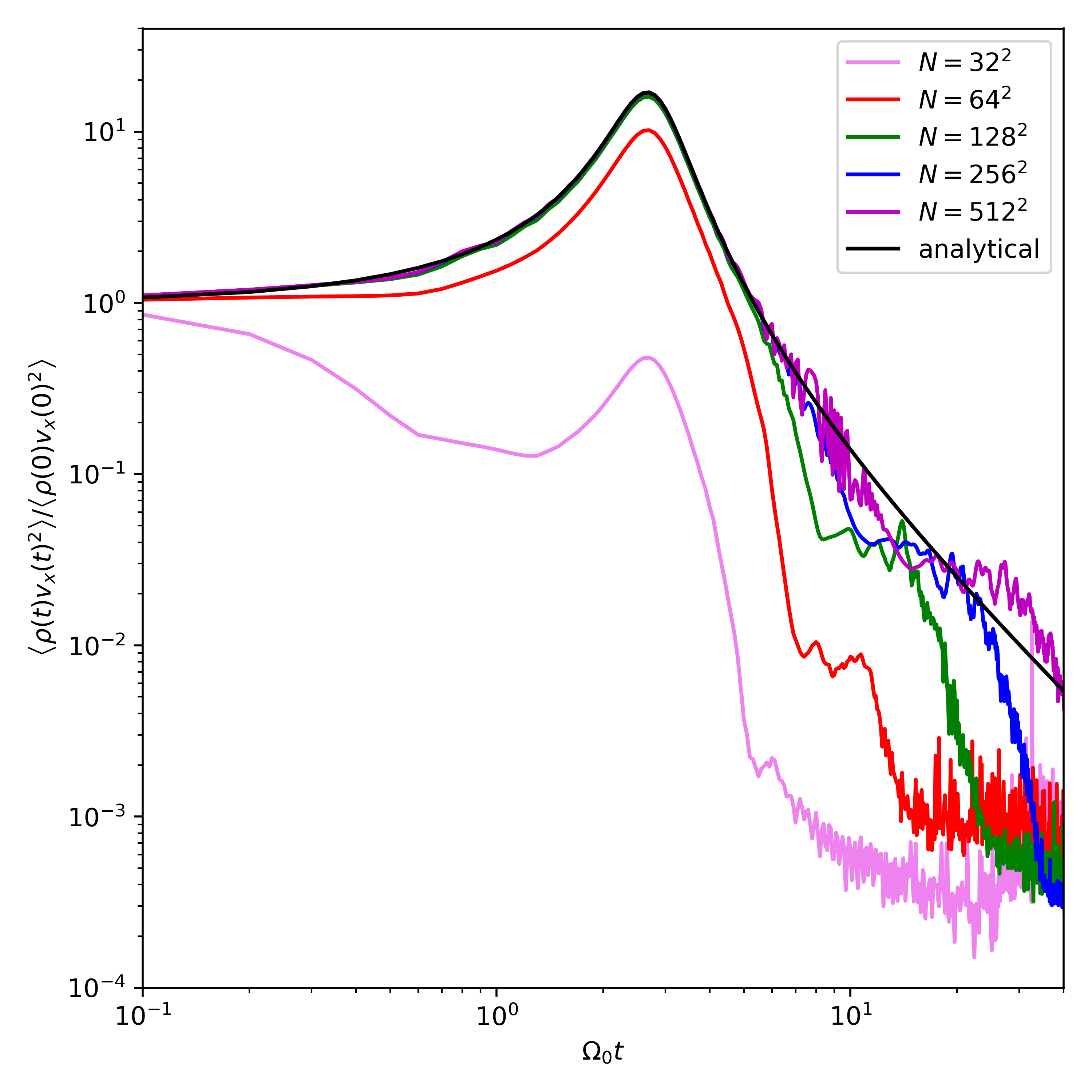}
    \caption{The evolution of the total kinetic energy associated with the movement in the $x$-direction for an incompressible hydrodynamic shearing wave.    The black line corresponds to the analytical solution of eqn.~(\ref{eq:HDShearingWaveEkinAnatlytical}) while the colored lines give measurements for simulations with different numerical resolutions, as labelled.}
    \label{fig:energyHDWaveEvolution}
\end{figure}

\begin{figure}
    \centering
    \includegraphics[width=1.0\linewidth]{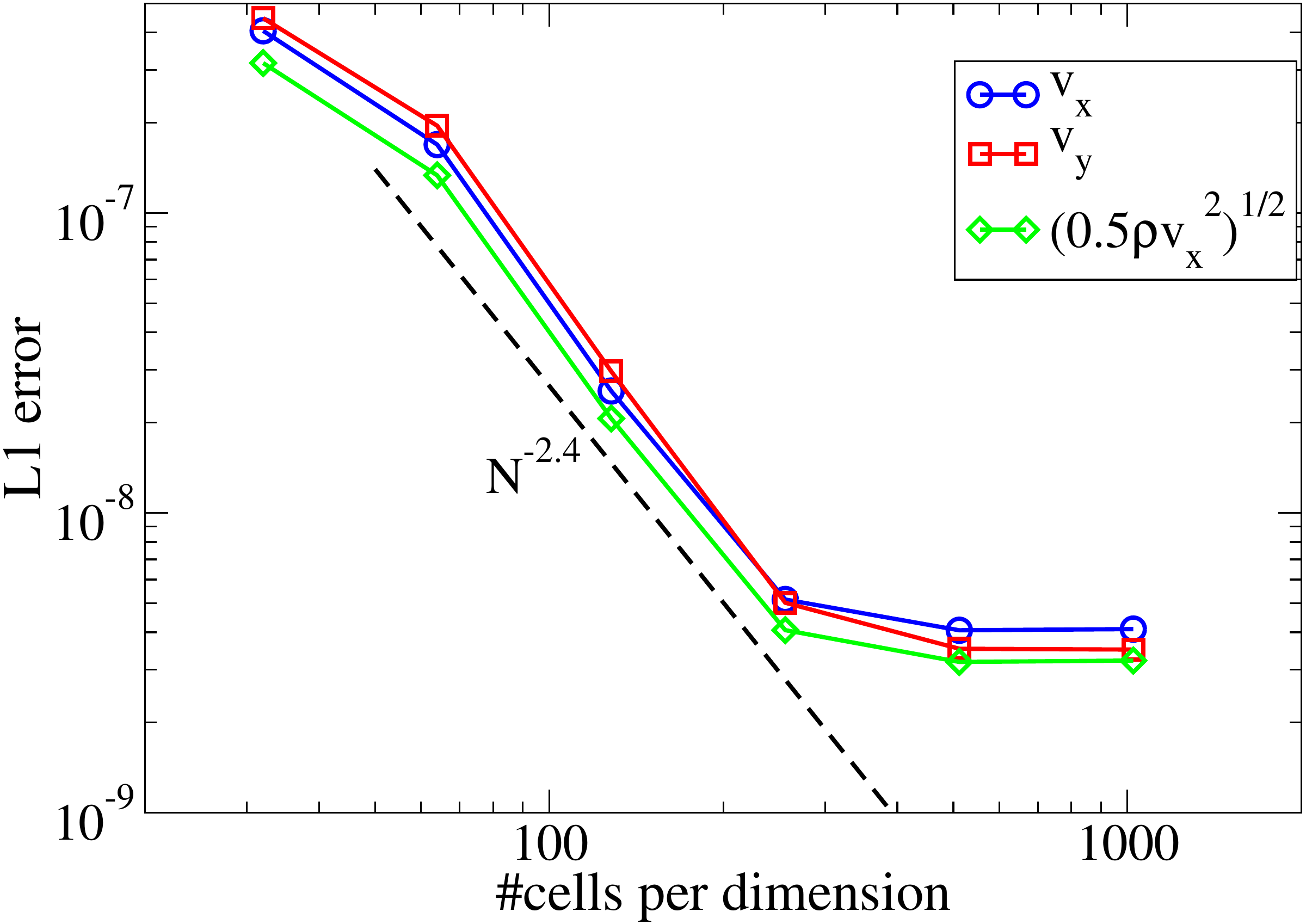}
    \caption{The $L_1$ error of the velocity and the kinetic energy density of the motion in the $x$-direction as a function of resolution for the simulation of an incompressible hydrodynamic shearing wave. The $L_1$ error was averaged over the time intervall $0 \leq  \Omega_0\,t < 6$. }
    \label{fig:ConvergenceHDWaveEvolution}
\end{figure}

In the second row of \cref{fig:groundStateL1Courant} we show the corresponding $L_1$ errors as a function of the Courant number now in the three dimensional case. We again find, as expected, $A_1 \approx 2$ for the second-order RK method, but now we recover a significantly better convergence rate than 2 for the third-order RK time integration. In the latter case, the maximum error decreases slower than the average errors, which means that for small time step sizes relatively large local errors dominate the average error budget. The fact that the third-order RK method does not decrease the error with full third order is a hint that the code likely still contains some inaccuracies that more prominently show up in two dimensions, and less so in three dimensions. One culprit  might lie in the estimate of the spatial velocity of an interface, which only takes into account the movement of the two adjacent cells and not the movement of other neighbouring cells, even though they introduce changes in the size of the facets.

Finally, we note again that we can completely suppress the errors in the time integration by starting with a perfectly structured mesh like a Cartesian or hexagonal grid, and only allow for a restricted mesh movement with $\vec{v}_{\rm mesh} = (0,-q \Omega_0 x,0)$ that is independent of time.  At all subsequent times, the mesh cells then still contain  symmetries that cancel truncation errors in the time integration, which typically leads to similar or even better results than for  the unstructured mesh cases. Note that this approach still removes the background bulk velocity from the calculation of the size of the time steps, it does have lower and spatially uniform advection errors compared to a Eulerian treatment, and the handling of the boundaries is manifestly translationally invariant. These properties are attractive, and thus such a scheme can be viewed as an interesting hybrid between a freely moving mesh and a static grid code. However, we will not discuss this scheme further in this paper, but it might nevertheless be of interest in cases where a very small level of numerical noise is required.

\begin{figure*}
    \centering
    \includegraphics[width=0.8\linewidth]{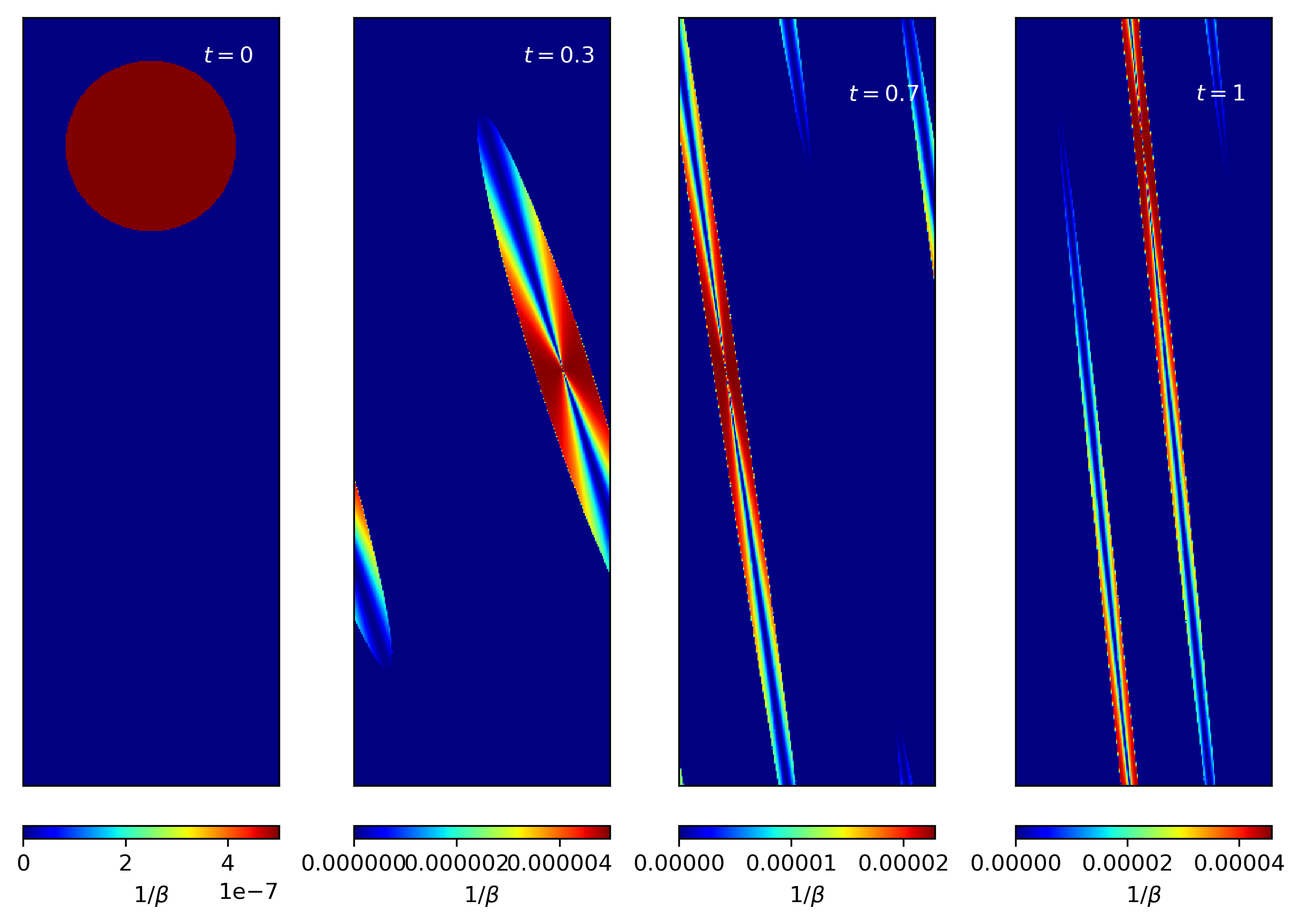}
    \caption{Strength of the magnetic field at different times in a simulation of the advection of a magnetic field loop in a shearing box.}
    \label{fig:advectionFieldLoop}
\end{figure*}

\subsection{Epicycle oscillations}
\label{sec:epicycleOsci}

Small perturbations to the velocity of the ground state of the shearing box do not get damped but instead lead to oscillations. In the case of $q=1.5$, they correspond to elliptical orbits in global disk simulations. If one assumes a perturbation $v_x = v_{x0}$ at $t=0$, the evolution of the velocities is given analytically by:
\begin{equation}
    v_x (t) = v_{x0}\cos(\Omega t),
\end{equation}
\begin{equation}
    v_y (t) = -v_{x0} \sqrt{\frac{2-q}{2} }  \sin(\Omega t).
\end{equation}
An important conserved quantity for these motions is the so-called epicycle energy 
\begin{equation}
    E_{epi} =  \frac{\left(v_x^2 + \frac{2}{2-q} v_y^2 \right)}{2} = {\rm const}.
    \label{eq:epicycleOsciEnergy}
\end{equation}
Obtaining this conservation numerically
requires special care in the implementation of the source terms of the shearing box. 

To test for this, we add a perturbation $v_{x0} = -0.1 c_s$ to the ground  state of a two-dimensional shearing box with $\Omega = 1.5$ and $c_s = 10^{-3}$.  We use a box size $L_x = L_y = 10$ and start with a randomly perturbed Cartesian grid. In \cref{fig:EpicycleAmplitude10} we show that even for the low resolution of $10^2$ cells the errors in the epicycle energy do not grow with time. However, if we do not add the predictor step for the second application of the source terms, as described in Section~\ref{subsec:sourceTermsImplementation}, the epicycle energy grows exponentially. In \cref{fig:EpicycleCongergence}, we show that the errors in the velocity as well as in the epicycle energy decrease with second order in the spatial resolution, confirming that epicycle oscillations are accurately treated by our shearing box implementation.

\subsection{Evolution of a hydrodynamic shearing wave}

To test the implementation of our boundary conditions for non-axisymmetric conditions we simulate the evolution of a hydrodynamic shearing wave \citep{johnson2005linear, balbus2006exact, shen2006three}. The setup follows \cite{stone2010implementation}. We use a two-dimensional box of size $L_x = L_y =1$, an isothermal equation of state with sound speed $c_s = 1.29 \times 10^{-3}$, $\Omega_0 = 10^{-3}$ and initial wave vector $2 \pi \bm k_0 / L_x = (-8,2)$. We choose  initial amplitudes $v_{x,0} = 10^{-4} c_s$ and $v_{y,0} = -k_{x,0} /k_{y,0} v_{x,0}$, and a constant background density $\rho = 1$.

The shearing box leads to a time-dependence of the wave \citep{shen2006three}:
\begin{subequations}
\begin{equation}
    k_x(t) = k_{x0} + q  \Omega_0  k_{y0} t,
\end{equation}
\begin{equation}
    v_x(t) = v_{x,0} \sqrt{\frac{k_{x,0}^2 + k_{y,0}^2}{k_x^2+k_{y,0}^2}} \cos(k_x(t) x + k_{y,0} y),
\end{equation}
\begin{equation}
       v_y(t) = - v_{x,0} \sqrt{\frac{k_{x,0}^2 + k_{y,0}^2}{k_x^2+k_{y,0}^2}} \frac{k_x(t)}{k_y} \cos(k_x(t) x + k_{y,0} y) .
\end{equation}
\end{subequations}
The kinetic energy gets amplified and reaches a maximum at $t= 8 / (3\Omega_0)$. In \cref{fig:energyHDWaveEvolution}, we compare the evolution of the total kinetic energy in the $x$-direction for different resolutions with the analytical value
\begin{equation}
 E_{{\rm kin},x} = \int_{0}^1 {\rm d}x \int_0^1 {\rm d}y \frac{1}{2} \rho v_x^2 = \frac{\rho v_{x,0}^2}{4}\frac{k_{x,0}^2 + k_{y,0}^2}{k_x^2+k_{y,0}^2}.
 \label{eq:HDShearingWaveEkinAnatlytical}
\end{equation}

For high resolution ($N=128^2$ and higher), we accurately recover the amplification of the kinetic energy. At later time, the kinetic energy decreases in our simulations due to numerical viscosity, but this effect becomes smaller if we increase the resolution, as expected. If there are only 4 cells per initial wavelength, the numerical viscosity starts to noticeably damp the wave already at the beginning. 

In \cref{fig:ConvergenceHDWaveEvolution} we show the average $L_1$ error for different fluid quantities as a function of resolution in the time interval $0 < t < 6 /\Omega_0$. As discussed earlier, for low resolution the numerical viscosity is high and we find large deviations from the analytical solution and a slow convergence rate. We then find a regime with seemingly faster than second-order convergence for high resolution, while eventually the errors start to become constant. The latter issue could be explained by the fact that the analytical solution was derived under the assumption of incompressibility, which is not actually compatible with the small density variations observed in our simulations.

\subsection{Advection of a weak magnetic field loop}

As a first test for the implementation of magnetic fields in the shearing box, we evolve a dynamically unimportant magnetic field loop in a homogeneous medium in the presence of background shear. We use a box with size $L_x =3$ and $L_y = 8$, a background density $\rho = 1$, an isothermal equation of state with sound speed $c_s = 1$, and the parameters $\Omega = 1$ and $q=3/2$ for describing the shearing of the box. The field loop has a radius $r=1$ with magnetic strength $\beta = \frac{2 c_s^2 \rho_0}{B_0^2} = 2 \times 10^6$, so the magnetic pressure is very small compared to the thermal pressure. Besides the background shear flow, we add a velocity $v_x = c_s$ to seed an epicycle oscillation of the field loop. This allows us to also check the implementation of the boundary conditions.  We use a uniform resolution of $1800 \times 600$ cells and add random displacements of 1\% of the mean particle spacing to the mesh generating points. In \cref{fig:advectionFieldLoop}, we show the resulting $\beta$ at different times. One can see no traces of the boundary conditions, and the shape of the field loop is well preserved throughout the evolution.

\subsection{Evolution of a compressible magnetic shearing wave}

Another sensitive MHD test for our shearing box implementation is the evolution of a compressible magnetic shearing wave in three dimensions \citep{johnson2007magnetohydrodynamic, stone2010implementation}. We use a box of size $L_x = L_y = L_z = 0.5$, an initially Cartesian grid with a constant number of cells per dimension, and an isothermal equation of state with sound speed $c_s = 1$. As  initial conditions we specify
\begin{subequations}
\begin{equation}
    \rho(\bm x) =  1+  5.48082  10^{-6} \cos(\bm k_0 \cdot \bm x),
\end{equation}
\begin{equation}
   \bm v(\bm x) = \begin{pmatrix} -4.5856  \\ 2.29279\\ 2.29279 \end{pmatrix} 10^{-6} \cos(\bm k_0 \cdot \bm x) - \begin{pmatrix}0\\ q \Omega_0 x\\0 \end{pmatrix},
\end{equation}
\begin{equation}
   \bm B(\bm x)/\sqrt{4\pi} = \begin{pmatrix} 5.48082  \\ 10.962\\ 0 \end{pmatrix} 10^{-7} \cos(\bm k_0 \cdot \bm x) + \begin{pmatrix}0.1 \\ 0.2\\0\\ \end{pmatrix},
\end{equation}
\end{subequations}
with a wavevector $\bm k_0 = (-2,1,1) \times 2\pi / L$.

\begin{figure}
    \centering
    \includegraphics[width=1\linewidth]{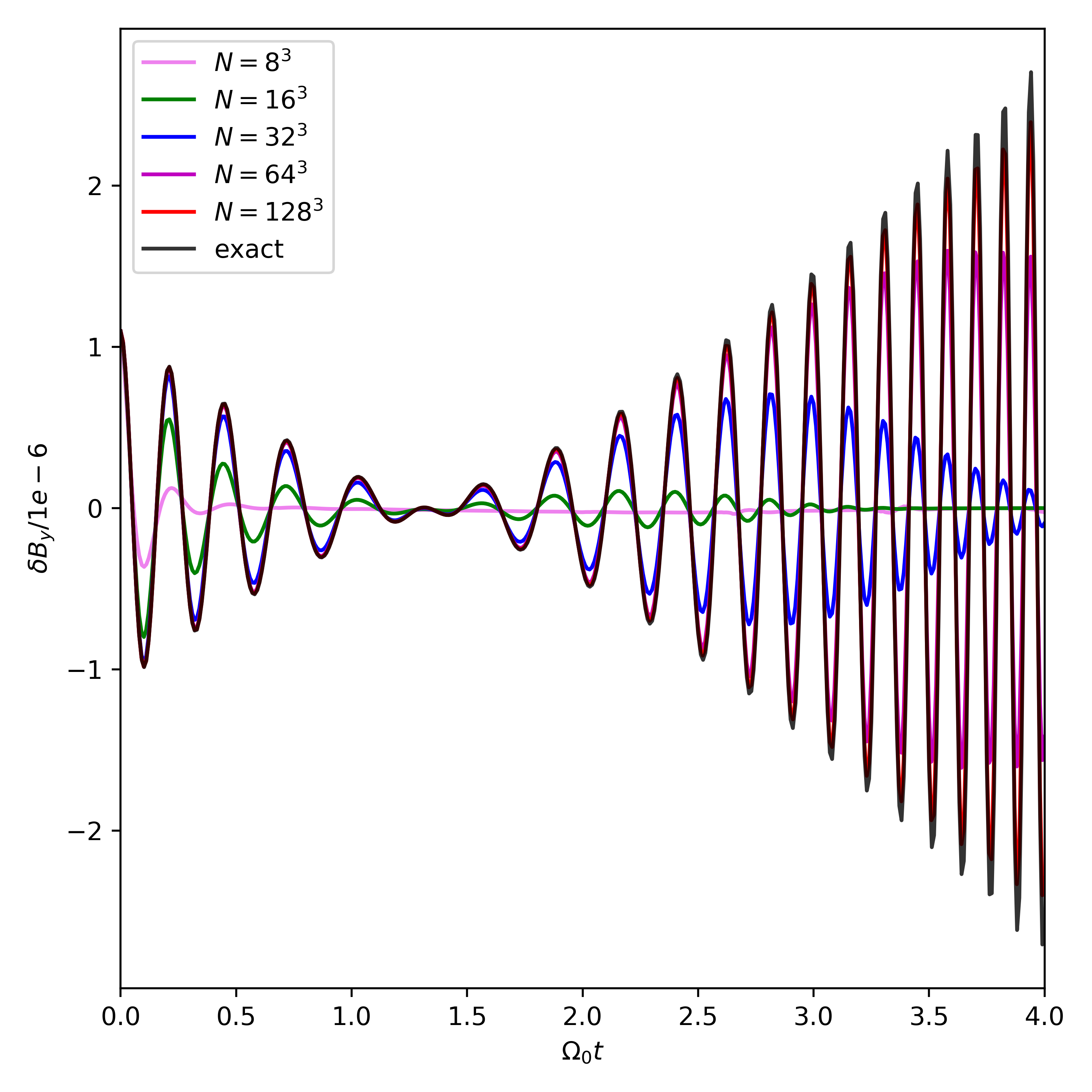}
    \caption{Evolution of the amplitude of the magnetic field $\delta B_y$ as a function of time in a compressible MHD shearing wave test, for different numerical resolutions, as labelled. }
    \label{fig:MHDShearingWavedBy}
\end{figure}

Due to shearing motion, the wavevector  $\bm k (t) = \bm k_0 +(q \Omega_0 t,0,0) \times 2\pi /L$ as well as the $y$-component of the average magnetic field $\hat{B}_y(t)/\sqrt{4 \pi} =  0.2 - 0.1 q \Omega_0 t$ evolve as a function of time. To obtain the evolution of the amplitudes of the wave we integrate the system of differential equations  given in \citet[][their eqns. 13--16]{johnson2007magnetohydrodynamic}. In \cref{fig:MHDShearingWavedBy} we compare the real part of the amplitude of the $y$-component of the magnetic field in our simulations, defined as
\begin{equation}
    \delta B_y = 2 \int_V (B_y - \hat{B}_y) \cos[\bm k(t) \bm x] {\rm d}\bm x,
\end{equation}
with the analytic values, as a function of the employed numerical resolution. 

\begin{figure}
    \centering
    \includegraphics[width=1.0\linewidth]{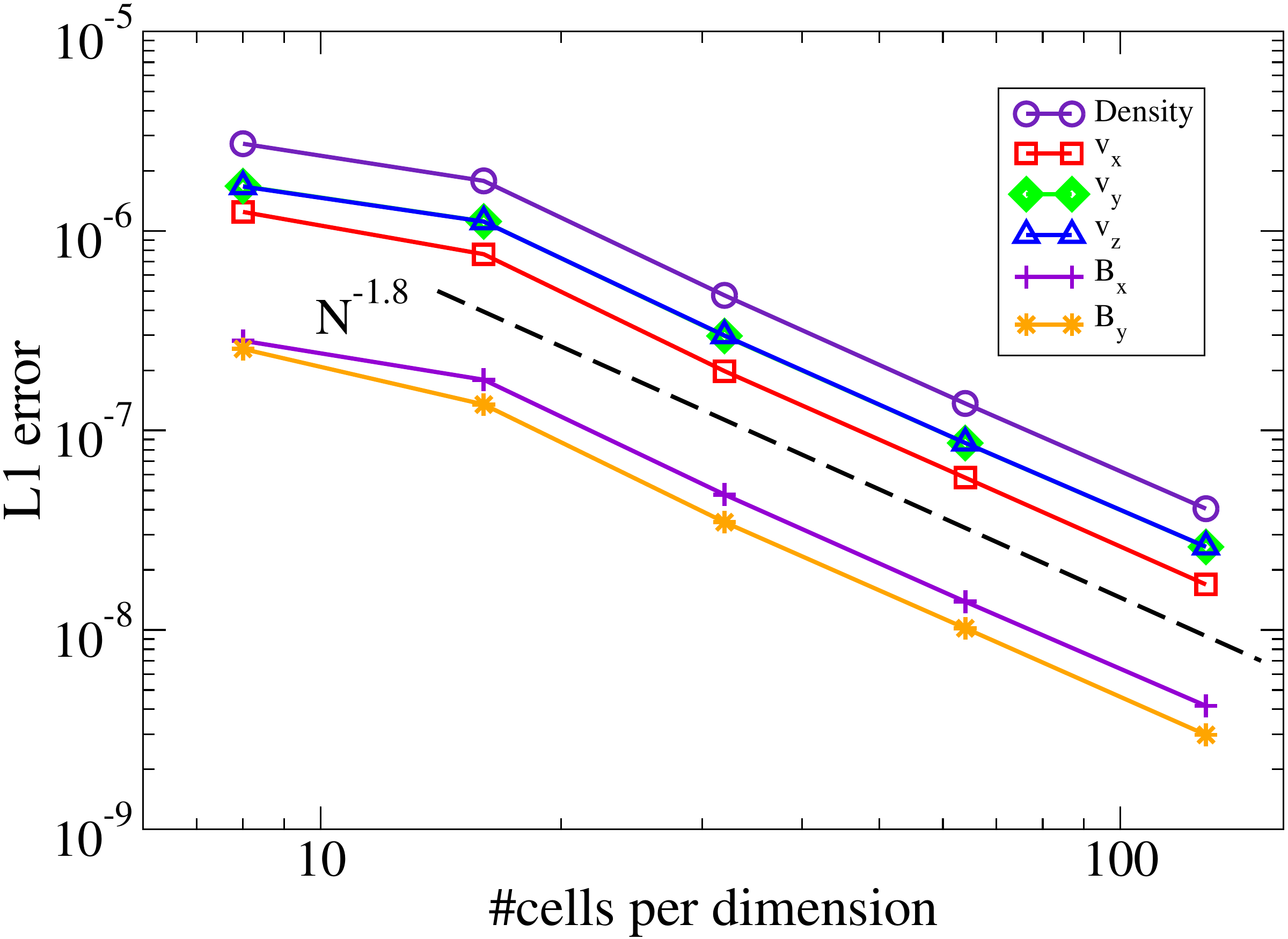}
    \caption{The $L_1$ error norm for different fluid quantities averaged over the time period $0< \Omega_0\, t < 2$  in a compressible MHD shearing wave test. The black dashed line shows the slope of $-2$ that we would ideally expect from our code, while the actual measurements as a function of resolution are shown by colored lines, as labelled.}
    \label{fig:MHDShearingWavedConvergence}
\end{figure}

The wave gets damped by numerical viscosity and resistivity. Both dissipative effects decrease with increasing resolution, allowing us to follow the evolution of the wave for longer times if the resolution is higher. To more quantitatively analyze the convergence properties of our code, we show in \cref{fig:MHDShearingWavedConvergence} the $L_1$ error of different quantities averaged over the time period $0< \Omega_0\, t< 2$, as a function of resolution. For sufficiently high resolution, we find a close to second-order convergence in all quantities.

\begin{figure*}
    \centering
    \includegraphics[width=1\linewidth]{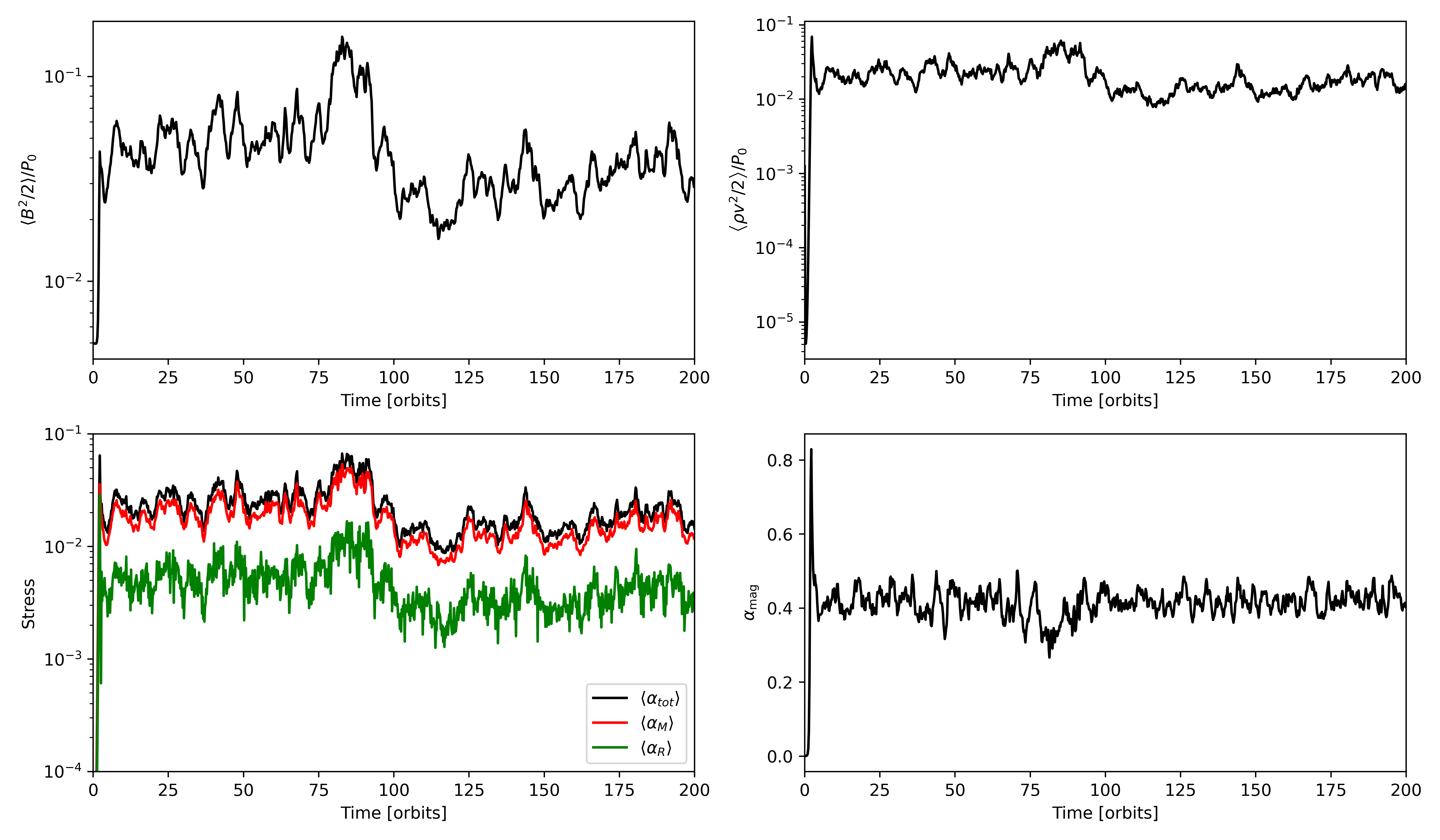}
    \caption{The temporal evolution of the average magnetic energy density (top left), kinetic energy density (top right), Maxwell and Reynolds stresses (bottom left), and of the normalized magnetic stress (bottom right) for a simulation of the magneto-rotational instability (MRI) without mean field and using standard Dedner divergence cleaning (i.e.~$c_{h0} = 1$). The simulation method shows a sustained quasi-stationary MRI even at moderate numerical resolution. } 
    \label{fig:MRI_data}
\end{figure*}

\begin{table*}
    \centering
    \begin{tabular}{c|c|c|c|c|c}
    \hline
        Origin sim. & 100 $\left<\alpha_{\rm tot}\right>$ &100 $\left<\alpha_M\right>$ & 100$\left<\alpha_R\right>$  & $\left< B^2 /2 \right> /P_0$ & $\left< \rho v^2 /2  \right> /P_0$ \\
        \hline
        \cite{shi2016saturation} & 5.16 & 4.17 & 0.99& 0.1301 & 0.0417\\
        This work ($c_{h0} = 0.1$) & 3.21 & 2.56 & 0.65 & 0.066 & 0.027\\
        This work ($c_{h0} = 0.2$) & 3.59 & 2.89 & 0.71 & 0.069 & 0.032\\
        This work ($c_{h0} = 0.5$) & 4.16 & 3.34 & 0.82 & 0.093 & 0.035\\
                 This work ($c_{h0} = 1$) & 2.20 & 1.74 & 0.46 & 0.043 & 0.020  \\
        This work ($c_{h0} = 2$)& 1.04 & 0.81 & 0.23 & 0.021 & 0.010\\
        This work ($c_{h0} = 5$)& 0.42 & 0.33 & 0.09 & 0.011 & 0.005\\
        \hline
    \end{tabular}
    \caption{A comparison of different time-averaged fluid quantities in simulations of the MRI using a tall box without net field. We give our results for a number of different settings of the Dedner cleaning speed ($c_{h0}$) and compare to the results from \protect\cite{shi2016saturation} (their simulation x1y4z4r32). All simulations use 32 cells per scale height and are based on identical initial conditions. The fluid quantities are averaged over 150 orbits, starting after the first 50 orbits.}
    \label{tab:MRI_time_average}
\end{table*}

\section{Nonlinear magneto-rotational instability without net field}
\label{sec:nonlinearMRI}

Finally, as a first example for an application of our new shearing box implementation to an interesting nonlinear problem we consider a simulation of the magneto-rotational instability (MRI), here without a mean magnetitic field and in an unstratified box. In a standard box size ($L_x = L_z = 1\,H$) previous studies found a non-converged behaviour of the MRI \citep{fromang2007mhd, fromang2007mhd2}. We therefore adopt a tall box set-up ($L_y = L_z = 4H = 4 L_x$) as in \cite{shi2016saturation} who achieved with this configuration convergence with the {\small ATHENA} code \citep{stone2008athena, stone2010implementation}. This test was also used in \cite{deng2019local} to show that with a resolution of 48 particles per scale height $H=c_s /\Omega_0$, the {\small GIZMO} code \citep{hopkins2015new} was not able to sustain the MRI for more than 20 orbits.  

We use a comparatively low resolution of 32 cells per scale height, an isothermal equation of state with $c_s =1$, and the shearing box parameters $q=3/2$ and $\Omega_0 = 1$. For the initial conditions, we use the ground state of the shearing box (\ref{eq:groundState}) with constant density $\rho = 1$, a purely vertical magnetic field
\begin{equation}
   \bm B= B_0 \sin \left(2\pi x/L_x\right) \hat{e}_z,
\end{equation}
and we add uniformly distributed random noise of maximum amplitude $0.05\, c_s$ to the initial velocities of all cells in order to seed the MRI. The initial field strength $B_0$ is defined by the parameter  $\beta = {2 c_s^2 \rho_0}/{B_0^2}= 400$, and we use the Dedner approach for divergence cleaning. We vary the strength of the cleaning by multiplying the standard variable $c_h$ in the Dedner scheme with a constant prefactor $c_{h0}$.  Smaller values mean weaker cleaning, and an effectively lower numerical resistivity. 

To analyse the simulations we define the volume weighted average of a quantity $X$ as
\begin{equation}
\left <X \right > = \frac{\int X {\rm d}V}{\int {\rm d}V},    
\end{equation}
as well as the Maxwell stress
\begin{equation}
\alpha_{M} = - \frac{B_x B_y }{ P_0} ,
\end{equation}
and the Reynolds stress
\begin{equation}
\alpha_{R} =  \frac{\rho v_x (v_y- \overline{v_y}) }{ P_0},
\end{equation}
where $\overline{v_y}$ denotes the background mean flow and $P_0 = 1$ is the initial pressure. The saturation of the MRI can be measured in terms of the relative Maxwell stress:
\begin{equation}
\alpha_{\rm mag} = - \frac{\left <B_x B_y \right >}{ \left < B^2 \right >},
\end{equation}
which goes to 0 if the MRI dies out.

In \cref{fig:MRI_data} we show the evolution of different fluid quantities relevant for the MRI. After an initial growth of the magnetic energy, we find as expected a saturated state. In Table~\ref{tab:MRI_time_average} we compare several time-averaged quantities with the equivalent simulations form \cite{shi2016saturation}. The measured stresses and energies are all slightly smaller than in {\small ATHENA}, which hints that the Dedner cleaning is more diffusive than the constrained transport scheme used in \cite{shi2016saturation}. For $c_{h0} = 0.5$ we find the strongest stresses in the saturated phase. For higher $c_{h0}$, the stresses decrease due to a larger numerical resistivity.

Overall, the results are very positive as they are superior compared to previous studies of the MRI with Lagrangian techniques, and are nearly as good as those obtained with a Eulerian constrained transport method. Given that our method uses  a fully dynamic and unstructured mesh, which does not benefit from fortuitous cancellations of truncation errors, this encourages the use of the new method for more complicated physics applications. In a future study we for example plan to further analyse this system by modifying the resolution, the cleaning speed $c_h$, and the initial magnetic field, and it will be especially interesting to study the case of a stratified shearing box as well. We will also further compare our results with the studies in \cite{deng2019local} and \cite{wissing2022magnetorotational} that both used Lagrangian techniques.

\section{Summary and conclusions}

In this work, we have focused on the technical realization of the shearing-box approximation in the moving-mesh code {\small AREPO}. The shearing-box is ideal to achieve very high local resolution in rotationally supported disk flows, much higher than can be readily  obtained in global disk simulations. This allows, for example,  a study of the non-linear behaviour of crucial fluid instabilities in such disks, such as the magneto-rotational instability.

While the mesh construction algorithm of {\small AREPO} can be straightforwardly generalized to cope with the special shear-periodic boundary conditions appearing in the radial direction, we found that the default version of the {\small AREPO} code produces an uncomfortably high level of `grid-noise' in the ground-state flow of the shearing box. In previous applications of {\small AREPO}, it had often been noted that the code's accuracy was reduced in situations with strong shear, because here mesh faces turn rapidly, and the distortion and time variation of the geometry of mesh cells are particularly strong. In this sense, the shearing-box is the worst situation one can possibly imagine, because shear appears {\em everywhere}, and {\em all the time}. In this paper, we  were thus compelled  to identify the origin of these inaccuracies in a clearer way and in more detail than had been understood before. But this then led also the way to overcome these source of noise and develop a significant improvement of the accuracy of the integration scheme of {\small AREPO}. This advance is of particular importance for cold, strongly shearing flows, but it ultimately is also beneficial for all other applications of the code.

In particular, we found that for general unstructured Voronoi meshes it can be necessary to integrate the fluxes over mesh faces with more than one Gauss point in order to reduce truncation errors to a level that is achieved by symmetric cells in part through cancellation effects on opposite sides of cells. We have also found that a minor change in the slope limiting scheme of {\small AREPO} is helpful to avoid that the slope limiting can be needlessly triggered in situations with highly distorted cell geometries. Finally, we rectified a minor inconsistency in the time integration scheme when mesh regularization motions are applied by the code. We note that to identify these improvements, examining the ground state of the shearing box in detail, a seemingly trivial state, proved essential. 

With the newly proposed improvements in place, we have shown that {\small AREPO} is able to very accurately simulate the shearing-box. Also, our improvements are beneficial for other types of simulations where strong shear is present, such as the Yee vortex. An advantage compared to Eulerian schemes is that the local truncation error of our approach is fully uniform within the shearing box, and does not increase towards the radial box boundaries as in Eulerian methods. In fact, the calculation is fully translationally invariant. When replicating multiple copies of the primary simulation domain using the (shear) periodic boundary conditions, and sticking them together at the boundaries to cover a larger domain, one could not tell any more where the original box boundary had been.

As a first important application, we have considered the magneto-rotational instability without a background field. {\small AREPO} is able to sustain the MRI even at comparatively low resolution, unlike the Lagrangian MFM approach. Our results show evidence for a somewhat higher numerical resistivity compared to the constrained transport method used by the {\small ATHENA} code. Perhaps different methods for magnetic divergence control can improve on this in the future, but already now {\small AREPO} should be a versatile and flexible tool for studying disks with the shearing-box approximation, in particular due to its ability to seamlessly increase the local resolution in situations with fragmentation and local (gravitational) collapse. We will consider such physics problems in forthcoming applications of the code.

\label{sec:discussion}

\section*{Acknowledgements}
We thank the anonymous referee for an insightful and constructive report that helped to improve the paper.
The authors acknowledge helpful discussions with R\"udiger Pakmor.

\section*{Data Availability}
The data underlying this paper will be shared upon reasonable request to the corresponding author.

\bibliographystyle{mnras}
\bibliography{main.bib}

\begin{appendix}

\section{Gauss-Legendre integration}

The class of Gauss-Legendre quadrature rules numerically integrates functions $f(\bm x)$ as a weighted sum of the function evaluations $f(\bm x_i)$ at different $\bm x_i$.
\subsection{One dimensional integrals}
\label{subsec:gauss2Dims}
In one dimension, any integral over a finite interval can be transformed into the form (\ref{eq:1dGaussLegendre}). In Table~\ref{tab:GaussLegendre1D} we give the corresponding evaluation points $x_i$ and weights $w_i$ up to order $P=5$.

\begin{table}
    \centering
    \begin{tabular}{c|c|c}
        \hline
        P & $x_i$ & $w_i$ \\
        \hline
        \hline
        1 & 0 & 2\\
        \hline
        3 & $-1/\sqrt{3}$ & 1\\
        & $ 1/\sqrt{3}$ & 1\\
        \hline
        5 & $-\sqrt{3/5}$ & $5/9$\\
        & $0$ & $8/9$\\
        & $\sqrt{3/5}$ & $5/9$\\
            \hline
\end{tabular}
    \caption{The maximum order $P$ of the polynomials that can be integrated exactly with the corresponding evaluation points $x_i$ and weighting factors $w_i$.}
    \label{tab:GaussLegendre1D}
\end{table}

\subsection{Two dimensional integrals}

In two dimensions, we are integrating over polygons. There exists no general equation for polygons with more than four edges that could conveniently give us the minimum number of evaluation points needed to integrate a polynomial of order $N > 1$ exactly. For the unit square $[-1,1]\times [-1,1]$ one can construct such rules as  tensor product of  one-dimensional Gauss-Legendre integrations. But the coordinate transformation from a general quadrilateral to the unit square is in general nonlinear, which means that by transforming the original polynomial to the natural variables of the unit square, one also increases the order of the polynomial. This would in turn require a higher-order rule for the unit square, increasing the computational cost. We therefore choose a different approach and split the original polygon into triangles.  Integrals of polynomials over any triangle can be transformed into integrals of polynomials of the same order over the unit triangle defined by the local coordinates $0 \leq r \leq 1$, $0 \leq  s \leq 1-r$.

The integration rule then takes the form:
\begin{equation}
    I = \int_{0}^1 \int_0^{1-r} f(r,s)\, {\rm d}s\, {\rm d}r \approx \sum_{i=1}^M f(r_i, s_i) w_i.
\end{equation}
As in the one-dimensional case, the integral gets approximated by a weighted sum of function evaluations $f$ at different points $(r_i, s_i)$. In Table~\ref{tab:GaussLegendre2D} we give the coordinates and weights of those points that can be used to integrate polynomials of order $\leq P$ exactly. For a polynomial of order 3 (e.g.~the velocity flux function or the energy flux function for constant density) we require 4 evaluations.

\begin{table*}
    \centering
    \begin{tabular}{c|c|c|c|c|c}
            \hline
        P & N& $L_1$ & $L_2$& $L_3$&  $w_i$ \\
        \hline
        \hline
        1 & 1& $1/3$& $1/3$ &  $1/3$& 1\\
        \hline
        2 & 3 & $2/3$& $1/6$ &  $1/6$& $1/3$\\
        \hline
        3 & 1& $1/3$& $1/3$ &  $1/3$& $-9/16$\\
        & 3 & $3/5$& $1/5$ &  $1/5$& $25/48$\\
        \hline
        4 & 3 & $0.108103018168070$ & $0.445948490915965$ & $0.445948490915965$ & $0.223381589678011$ \\
        &  3 & $0.816847572980459$ & $0.091576213509771$ & $0.091576213509771$ & $0.109951743655322$ \\
                \hline
    \end{tabular}
    \caption{The maximum order $P$ of the polynomials over the unit triangle that can be integrated exactly with the corresponding evaluation points and weighting factors $w_i$.
    The two coordinates of the evaluation points can be constructed by $N$ cyclic uses of $L_1$, $L_2$ and $L_3$.
    The table values are adopted from \protect\cite{akin2005finite}.}
    \label{tab:GaussLegendre2D}
\end{table*}
\label{subsec:gauss3Dims}

\section{Details of test problem setups}
\subsection{Yee vortex}
\label{subsec:setupYee}
For the setup of the Yee vortex, we follow \cite{yee2000entropy} and \cite{pakmor2016improving}.
The mesh covers the interval $[-5,5] \times [-5,5]$, and we set the density $\rho$, velocity $\bm v$ and specific energy $u$ of each cell to the value of the continuous fields at the centres of mass of the cells. The continuous initial conditions are given at point $\bm r=(x,y)$ by:
\begin{subequations}
\begin{equation}
    T(\bm r) = T_{\rm inf} - \frac{(\gamma -1)}{8 \gamma \pi^2} e^{1-r^2},
\end{equation}
\begin{equation}
    \rho (\bm r) = T^{\frac{1}{1 - \gamma}},
\end{equation}
\begin{equation}
    v_x(\bm r) = -y \frac{\beta}{2 \pi} e^{\frac{1-r^2}{2}},
\end{equation}
\begin{equation}
    v_y(\bm r) = x \frac{\beta}{2 \pi} e^{\frac{1-r^2}{2}},
\end{equation}
\begin{equation}
    u(\bm r) = \frac{T}{\gamma -1}.
\end{equation}
\end{subequations}
We use the parameters $T_{\rm inf} = 1$, $\gamma = 1.4$ and $\beta = 5$.

For the setup of the Cartesian grid we use cells with side length $10/N$, where $N$ is the linear resolution. To setup a polar grid we use the method of \cite{pakmor2016improving}: We place the mesh-generating points in rings with size $d_{\rm ring} =  10/N$. We equidistantly add $2\pi r_{\rm ring} / d_{\rm ring}$ new mesh-generating points  in every ring at radius $r_{\rm ring} = i\, d_{\rm ring}$.  We only add points which lie in the simulation domain.

\subsection{Keplerian disk}

The setup is the same as in \cite{pakmor2016improving}.  We use a periodic box spanning the domain $[-2.5,2.5]\times [-2.5,2.5]$.  The disk itself is cold, which means the pressure forces are negligible compared to the gravitational force. As for the Yee-vortex, we again create a polar grid as initial condition and set the density $\rho$, the velocity $\bm v$ and the specific internal energy $u$ to the analytic values at the centres of mass of each cell. As background medium for $r = |\bm r| = |\left(x,y\right)| < 0.5$ and $r> 2$ we use
\label{subsec:setupKepler}
\begin{subequations}
\begin{equation}
    \rho = 10^{-5},
\end{equation}
\begin{equation}
    \bm v = 0,
\end{equation}
\begin{equation}
    u = \frac{5 \gamma}{2\rho} \times 10^{-5}.
\end{equation}
\end{subequations}
For the disk itself ($0.5 < r <2$) we use
\begin{subequations}
\begin{equation}
    \rho = 1,
\end{equation}
\begin{equation}
    \bm v(x,y) = \frac{1}{r^{3/2}}\begin{pmatrix}
    -y\\x
    \end{pmatrix} ,
\end{equation}
\begin{equation}
    u = \frac{5 \gamma}{2\rho} \times 10^{-5}.
\end{equation}
\end{subequations}
Here $\gamma = 5/3$ is the adiabatic index. As gravitational acceleration we use
\begin{equation}
    \bm g = -\frac{\bm r}{r \left(r^2+\epsilon^2 \right)},
\end{equation}
with $\epsilon = 0.25$ for $r < 0.25$, and $\epsilon = 0$ everywhere else.
\end{appendix}

\bsp	
\label{lastpage}
\end{document}